\documentclass[review]{elsarticle}

\usepackage{lineno}
\usepackage{amsmath}
\usepackage{multirow}
\usepackage{subcaption}
\usepackage{relsize}
\usepackage{algorithm}
\usepackage[noend]{algpseudocode}
\usepackage{listings}
\usepackage{slashbox}

\modulolinenumbers[5]
\newcommand{\norm}[1]{\left\lVert#1\right\rVert}
\journal{Journal of Speech Communication}

\usepackage{color}
\usepackage{verbatim}

\begin{document}

\begin{frontmatter}

%\title{On the use of asymmetric context windows \\ for reverberant speech recognition}
%\title{Asymmetric context windows for \\ DNN-based recognition of reverberant speech}
\title{Automatic context window composition \\ for distant speech recognition}
%\title{Feeding DNNs with asymmetric context windows \\ for counteracting reverberation}
\author{Mirco Ravanelli, Maurizio Omologo}
\address{Fondazione Bruno Kessler, Trento, Italy}

\begin{abstract}
Distant speech recognition is being revolutionized by deep
learning, that has contributed to significantly outperform previous
HMM-GMM systems. A key aspect behind the rapid rise and success of DNNs is their ability to better manage large time contexts.
With this regard, asymmetric context windows that embed more past than future frames have been recently used with feed-forward neural networks. This context configuration turns out to be useful not only to address low-latency speech recognition, but also to boost the recognition performance under reverberant conditions. 

%Previous studies have dealt only superficially  with the this asymmetric windowing mechanism, since most of them usually report some performance improvements without investigating too much on the reasons why they happen.  Contrary,
%To goal of this work is to explain the motivations behind the effectiveness of this approach, proposing a deep analysis focused on the role played by asymmetric contexts to counteract the harmful effects of reverberation. 
This paper investigates on the mechanisms occurring inside DNNs, which lead to an effective application of asymmetric contexts.    
In particular, we propose a novel method for automatic context window composition based on a gradient analysis.
The experiments, performed with different acoustic environments, features, DNN architectures, microphone settings, and recognition tasks show that our simple and efficient strategy leads to a less redundant frame configuration, which makes DNN training more effective in reverberant scenarios. 
%Experimental results also demonstrate that the proposed method outperforms the traditional application of a symmetric context window under all the reverberant conditions that have been explored.
   
%Based on the emerged experimental evidence, we also propose a novel algorithm for automatic context window composition which exploits a gradient analysis over the input frames.
\end{abstract}


\end{frontmatter}

%\linenumbers

\section{Introduction}
\label{sec:intro}
Distant Speech Recognition (DSR) represents a fundamental technology towards flexible human-machine interfaces.
%The human voice is the most natural way to communicate. Therefore, building computers that understand speech represents a crucial step towards easy-to-use human-machine interfaces \cite{roberto}. Automatic Speech Recognition (ASR) has recently been applied in several fields, such as web-search, intelligent personal assistants, car control, radiological reporting, and is currently  used by millions of users worldwide. 
%Nevertheless, most state-of-the-art systems are still based on close-talking solutions, forcing the user to speak very close to a microphone-equipped device. 
%Although this approach usually leads to better performance, users  might prefer to relax the constraint of handling or wearing any device to access speech recognition services, requiring technologies able to better cope with distant-talking interaction even in challenging acoustic environments. 
There are indeed various real-life situations where DSR is more natural, convenient and attractive  than traditional close-talking speech recognition \cite{dasr}.
For instance, applications such as meeting transcriptions and smart TVs have been studied over the past decade in the context of the AMI/AMIDA \cite{ami} and the DICIT \cite{dicit_1} projects, respectively.
More recently,  speech-based domestic control gained a lot of attention \cite{vacher,isidoros}.
To this end, the EU DIRHA project developed voice-enabled automated home environments based on distant-speech interaction in different languages \cite{lrec,dirha_asru}. Concerning this application, innovative commercial products, such as Amazon Echo and Google Home, have recently been introduced in the market. 
Robotics, finally, represents another emerging scenario, where users can freely talk with distant mobile platforms.

Several efforts have been devoted in the last years to improve DSR technology, as witnessed by the great success of some international challenges such as CHiME \cite{chime3}, REVERB \cite{revch_short}, and ASpIRE \cite{aspire}. A major role in improving this technology is being played by deep learning \cite{Goodfellow-et-al-2016-Book,lideng,dnn_shared}, which has contributed to significantly outperform HMM-GMM speech recognizers.
Deep learning, in effect, has been rapidly evolving during the last years, progressively offering more powerful and robust techniques, including effective regularization methods \cite{dropout,batchnorm}, improved optimization algorithms \cite{adam}, as well as better architectures \cite{cnn1,tdnn2,lstm_highway,gru1}.

%Despite this rapid evolution, state-of-the-art systems are still far from reaching a satisfactory level of robustness to acoustic conditions characterized by considerable levels of non-stationary noise and acoustic reverberation \cite{adverse}, keeping DSR a very active research field \cite{dsr_summary,pawel2,hain,dnn_rev,dnn_rev2,dnn3,ravanelli_SLT,ravanelli_icassp}. 
%This work is in line with these efforts and proposes a novel and computationally inexpensive methodology to counteract the harmful effects of reverberation.
A key aspect behind the success of deep learning in speech recognition is the ability of modern DNNs to perform predictions based on a large time context. 
A valid architecture able to learn long and short-term dependencies is represented by Recurrent Neural Networks (RNNs) as Long Short-Term Memory (LSTM) \cite{lstm} or Gated Recurrent Units (GRUs) \cite{gru1,ravanelli_is17}.
%An interesting RNN-based approach has been recently provided by encoder-decoder architectures coupled with attention models \cite{attention1}, where a neural network is jointly trained to decide which input information to analyze at each decoding step.
In order to simultaneously manage both past and future time contexts, the most suitable solution would be bidirectional RNNs \cite{graves}, that turned out to automatically learn (through their recurrent connections) how to properly exploit the contextual information.  The price to be paid for automatically learning contexts from speech data is an increased computational complexity. LSTMs, for instance, are based on a rather complex cell design based on three multiplicative gates, which normally require much more computations at each time steps if compared to a simpler feed-forward NN.
Bidirectional RNNs, moreover, can generate a sequence of posterior probabilities only after processing the entire sentence. 
Both of these features often impair their use in real-time/low-latency applications. 

To circumvent this drawback, unidirectional RNNs or feed-forward DNNs can be used. For low-latency applications, feed-forward DNNs still represent the most preferable choice for many practical applications, as witnessed by the numerous studies in the recent literature on real-time ASR systems  operating on devices with low computational power \cite{acw1,small2,small3,small4,small5,online2}. In line with these recent efforts, this work considers the aforementioned scenario, targeting standard feed-forward DNNs.  

In the case of feed-forward DNNs, the input features are typically gathered into a symmetric context window (SCW), that observes the current frame along with the same number of surrounding past and future ones. %For speech recognition applications, a symmetric context window composed of the same number of past and future frames is normally employed.  
Nevertheless, an asymmetric context window (ACW) that integrates more past than future frames has gained popularity for real-time/low-latency recognition of close-talking speech \cite{online2,acw1,acw_ct,tdnn2}. Interestingly, some recent papers have also evidenced the effectiveness of its application with distant speech recognition \cite{acw_rev,ravanelli15}, though a deep analysis is missing concerning the  
%Previous works, however, showed the performance benefits deriving from asymmetric contexts, but they did not investigate in depth on the 
conditions under which this approach becomes convenient with reverberated speech. 

The goal of this work is to better understand these aspects, and to
propose a methodology to derive an optimal context window (CW) according to the characteristics of the DSR task. The proposed $AutoCW$ algorithm, tested on different  tasks, datasets, microphone configurations as well as on different acoustic environments, exploits a gradient analysis  performed at an early stage of the DNN training.  Its application significantly reduces the efforts needed to find an optimal context, while improving ASR performance provided by the use of standard SCWs.

The rest of the paper is organized as follows. Sec. \ref{sec:acw} analyzes the effects of ACW on speech signals, input features, and DNN gradients. Sec. \ref{sec:algo} describes the proposed $AutoCW$ algorithm for automatic context window composition. In Sec. \ref{sec:ct}, an overview of the adopted experimental setup is provided, while the ASR results are reported in Sec. \ref{sec:exp}. Finally, Sec. \ref{sec:conc} draws our conclusions.

\section{Asymmetric Context Window for Counteracting Reverberation}
\label{sec:acw}
% \begin{figure*}[t!]
% \centering
%   \includegraphics[scale=0.61,trim={0cm 0 0.8cm 0},clip]{pipeline.jpg}
%   \caption{End-to-end pipeline of a HMM-DNN distant speech recognizer. The system is fed with a distant-talking signal y[n] that derives from the the original close-talking speech x[n] (not observable directly).} 
% \label{fig:dnn}
% \end{figure*}

%To better introduce the motivations behind the use of the asymmetric context window, it is useful to recall the full pipeline considered in this work. The speech recognizer, depicted in Fig. \ref{fig:dnn}, corresponds to the system configuration typically employed in most DSR systems. All the system stages, ranging from the preprocessing of the distant speech $y[n]$ to the generation of the output sequence of words $\hat{w}$, will be discussed  in the following sub-sections.

\begin{figure}[t!]
\begin{subfigure}{0.50\textwidth}
\includegraphics[scale=0.44,trim={0cm 0cm 0cm 0cm},clip]{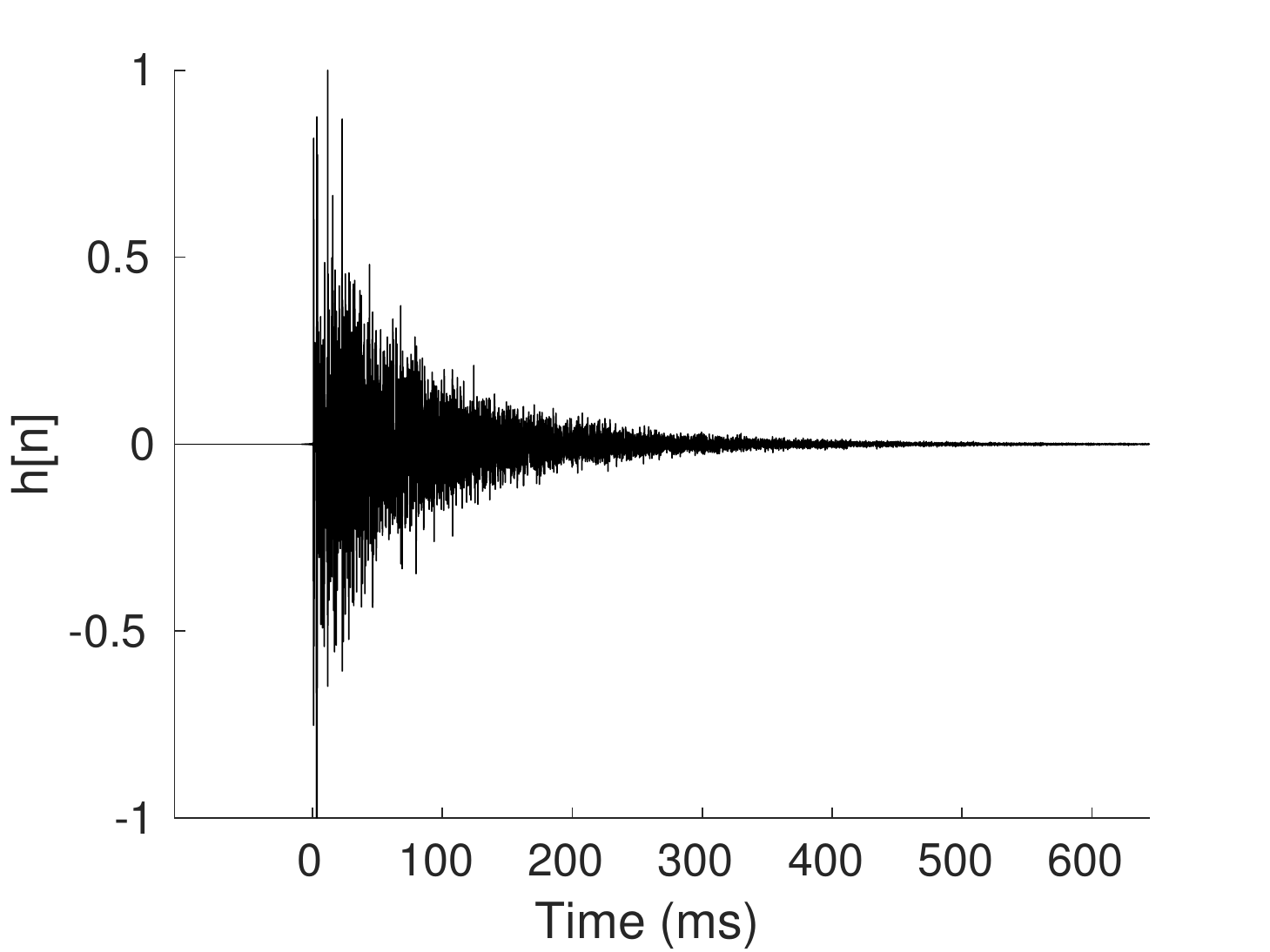}
\caption{Impulse Response $h[n]$}
\label{fig:ir_new}
\end{subfigure} \hspace{0.0\textwidth}
\begin{subfigure}{0.50\textwidth}
\includegraphics[scale=0.44]{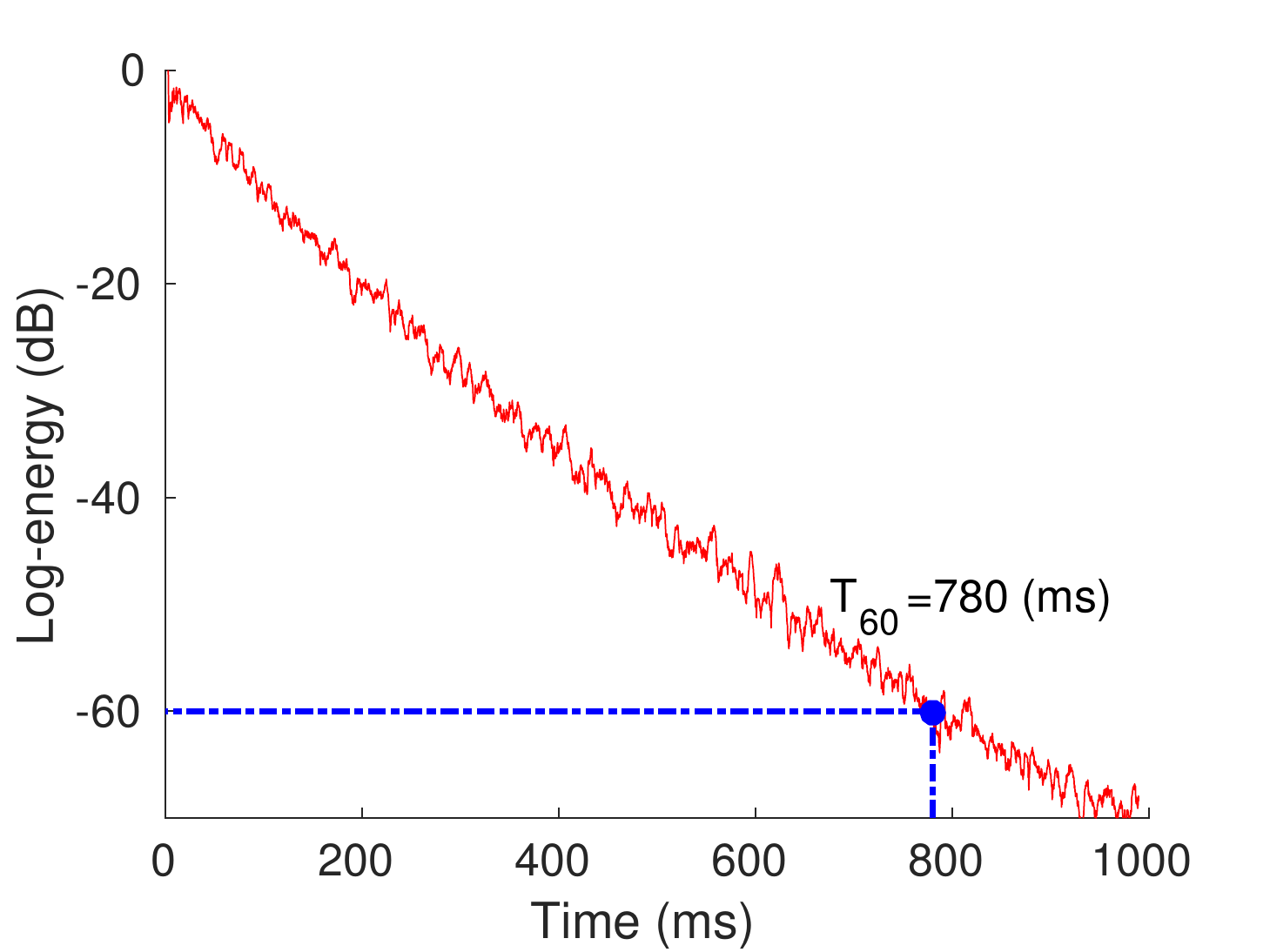}
\caption{Log-energy decay of $h[n]$}
\label{fig:ir_en}
\end{subfigure}
\caption{An impulse response measured in a domestic environment with a reverberation time of about 780 ms.}
\label{fig:ir}
\end{figure}

%\subsection{Environmental effects}
To better introduce the motivations behind the use of the ACW, it is useful to recall the effect of reverberation on a speech signal.
Let us describe a distant speech signal $y[n]$ by the following equation:
 \begin{equation}
 y[n]=x[n]*h[n]+v[n]
 \label{eq:cont}
 \end{equation}
where $x[n]$ is the close-talking signal (i.e, the speech signal before its propagation in the acoustic environment, which is assumed to be a latent variable not directly observed), $h[n]$ is the acoustic impulse response (IR) between source and microphone, and $v[n]$ is the additive noise introduced by the environment.
%{\color{blue} RIVEDERE IL SEGUENTE PARAGRAFO CHE POTREBBE ESSERE COMPATTATO}
The speech signal $x[n]$ is reflected many times 
%by the walls, the floor and the ceiling as well as by the objects within the acoustic environment. 
by walls, floor, and ceiling as well as by objects within the acoustic environment. 
Such a multi-path propagation, known as reverberation \cite{kutt}, is summarized by the IR $h[n]$, that can be modeled as a causal FIR filter (i.e., $h[n]=0 \quad \forall n<0$).
%which is typically characterized by the corresponding reverberation time $T_{60}$ \cite{Kutruff_o_altro}.
%In particular, if one assumes to deal with a linear time-invariant acoustical transmission system, the impulse response provides a complete description of the changes a sound signal undergoes when it travels from a particular position in space to a given microphone \cite{kutt}. %The impulse response $h[n]$ can be modeled as a long and causal FIR filter (i.e., $h[n]=0 \quad \forall n<0$), whose taps describe the propagation of the signal in the environment. 
%To estimate the amount of reverberation characterizing an acoustic enclosure, one of the most important measure is the reverberation time $T_{60}$, which is defined as the time required for a sound to decay 60 dB from its initial energy level. 
%Although $T_{60}$ varies significantly depending on the room acoustics, in standard environments it typically ranges between 250 ms and 850 ms. 
Fig. \ref{fig:ir_new} shows an IR measured in a living-room, whose log-energy decay (reported in Fig. \ref{fig:ir_en}) indicates that the reverberation time $T_{60}$ \cite{kutt} is about 780 ms.  

%For what concerns the additive noise component $v[n]$ of Eq. \ref{eq:cont},  it may be related to a variety of factors and classified into different possible categories, e.g., stationary vs unstationary, diffuse vs coherent, and it might result from the simultaneous combination of noise sources of different nature \cite{matassoni}. 

%and distributed in space. 

%As outlined in Sec. \ref{sec:intro}, one key advantage of deep neural networks is their ability to learn from very long observation windows. %This aspect is proved to be very important for speech recognition purposes and is one of the fundamental aspects that have favored deep learning over previous GMM/HMM systems.
%The speech sequence $x[n]$ propagates into the acoustic environment and is recorded by a far microphone. 
%\subsection{Feature extraction and frame concatenation}
In a DSR system,
the distant-talking signal $y[n]$
%captured by the distant microphone 
is processed by a feature extraction function $f(\cdot)$ that computes a sequence of feature frames
%\begin{equation}
%f(y[n])=\{ \mathbf{y_{1}},...,\mathbf{y_{k}},...,\mathbf{y_{N_{fr}}} \}
%\end{equation}
$f(y[n])=\{ \mathbf{y_{1}},...,\mathbf{y_{k}},...,\mathbf{y_{N_{fr}}}\}$,
where each frame $\mathbf{y_{k}}$ is a vector consisting of $N_{fea}$ features, and $N_{fr}$ is the total number of frames. Feature extraction is normally carried out by splitting the signal into small chunks (lasting 20-25 ms with an overlap of 10 ms), and by applying a transformation to each chunk. % which summarizes the speech sequence with a reduced set of  coefficients.
To broaden the time context the DSR system is not only fed with the current frame $\mathbf{y_{k}}$, but also with some surrounding ones. The set of frames feeding the system, i.e. the context window, is defined in the following way: 
%For each frame $\mathbf{y_{k}}$, a DNN is employed to perform frame-level phone predictions. In order to estimate more robust posteriors, the network is not only fed with the current frame $\mathbf{y_{k}}$, but also with some surrounding ones. The set of frames feeding the DNN, i.e. the context window, is defined in the following way: 
\begin{figure*}[t!]
\centering
  \includegraphics[scale=0.61,trim={0cm 0 0cm 0},clip]{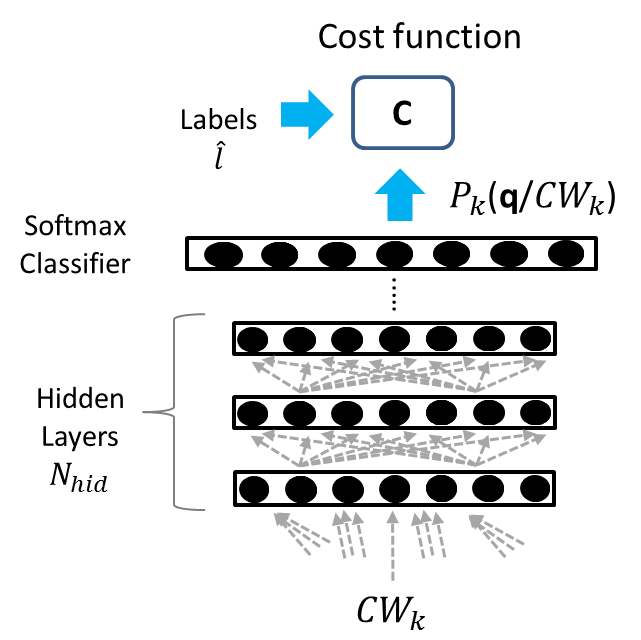}
  \caption{HMM-DNN pipeline used for hybrid speech recognition with feed-forward neural networks.} 
\label{fig:dnn}
\end{figure*}
\begin{equation}
CW_{k}=\{\mathbf{y_{k+p}}\} \quad \forall ~-N_{p}\leq p \leq N_{f} 
\label{eq:cw}
\end{equation}
where $N_{p}$ and $N_{f}$ are the number of past and future frames, respectively.
In standard SCWs $N_{p}=N_{f}$, while ACWs correspond to $N_{p}\neq N_{f}$. 
To account for different balance factors between past and future frames, let us introduce the coefficient $\rho_{cw}$ defined as follows:
\begin{equation}
\rho_{cw}(\%)=\frac{{N}_{p}}{N_{p}+N_{f}}\cdot 100
\end{equation}
%This factor will be 
It results that
$\rho_{cw}>50\%$ for an asymmetric context embedding more past than future frames, $\rho_{cw}=50\%$ for a symmetric context, and $\rho_{cw}<50\%$ when embedding more future frames.

The context window $CW_{k}$ then feeds a DNN, as depicted in Fig. \ref{fig:dnn}. The DNN processes the input features with several non-linear hidden layers and estimates a set of posterior probabilities $P_{k}(q \mid CW_{k})$.
The cost function $C(\hat{l},P_{k}(q \mid CW_{k}))$  optimized during training (e.g., cross-entropy) is computed from the reference labels $\hat{l}$ and the aforementioned predictions.

\subsection{Correlation Analysis}
%A  metric
A function that helps
study the redundancy introduced by reverberation is the cross-correlation. In particular, it is interesting to compute the cross-correlation $R_{xy}$ between the close-talking speech $x[n]$ and the corresponding distant-talking sequence $y[n]$. 
Let us assume that the additive noise $v[n]$ reported in Eq. \ref{eq:cont} is omitted here, to focus on reverberation only. 
%With this purpose,  we can expand the signal $y[n]$, defined in Eq. \ref{eq:cont},  in this way:
%\begin{equation}
%y[n] = \sum_{m=0}^{M-1} x[n-m]\cdot h[m]
%\end{equation}
%Note that, in order to focus our analysis on reverberation effects only, the additive noise $v[n]$ is omitted here.
%If we assume $x[n]$ to be a sequence of length L, and $h[n]$ to be an impulse response of length M, the cross-correlation $R_{xy}$ is defined as follows:
It can be easily shown that:
\begin{equation}
R_{xy}[n] =   \sum_{m=0}^{M-1} h[m] \cdot R_{xx}[n-m] 
\end{equation}
where $M$ and $R_{xx}[n]$ denote the finite length of the IR $h[n]$, and the autocorrelation of the close-talking signal, respectively.

%\begin{align}
%R_{xy}[n] & = \sum_{l=0}^{L-1} x[l]\cdot y[l+n] \\
%& = \sum_{l=0}^{L-1} x[l]\cdot \bigg( \sum_{m=0}^{M-1} x[l+n-m]\cdot h[m] \bigg) \nonumber \\
%& = \sum_{l=0}^{L-1} \sum_{m=0}^{M-1} x[l]\cdot x[l+n-m]\cdot h[m] \nonumber \\
%& = \sum_{m=0}^{M-1} h[m] \underbrace{\sum_{l=0}^{L-1} x[l]\cdot x[l+n-m]}_{R_{xx}[n-m]} \nonumber \\
%& =  \sum_{m=0}^{M-1} h[m] \cdot R_{xx}[n-m] \nonumber
%\end{align}

\begin{figure}[t!]
\begin{subfigure}{0.50\textwidth}
\includegraphics[scale=0.44,trim={0cm 0cm 0cm 0cm},clip]{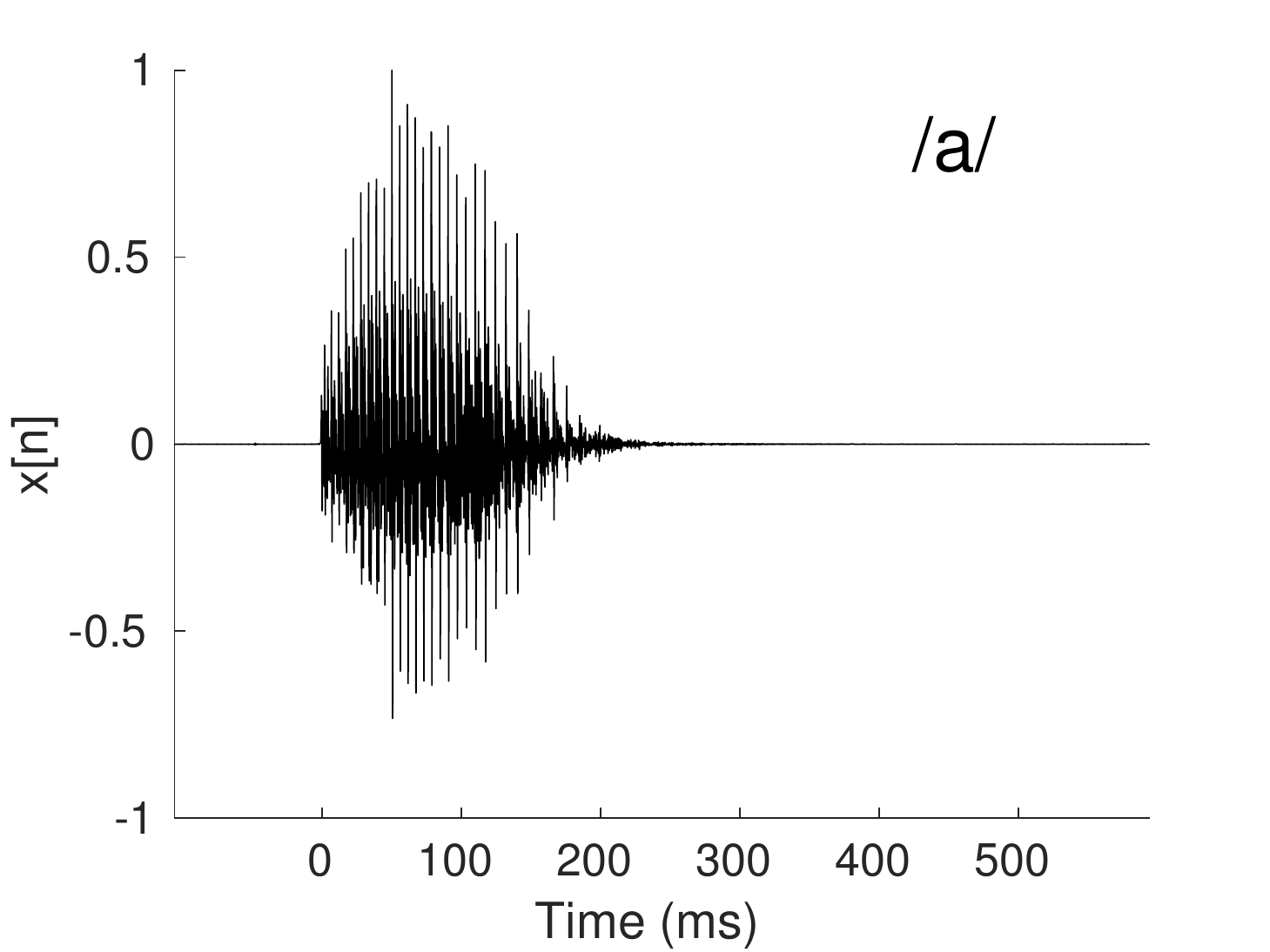}
\caption{Close-talking signal $x[n]$}
\label{fig:a}
\end{subfigure} \hspace{0.0\textwidth}
\begin{subfigure}{0.50\textwidth}
\includegraphics[scale=0.44,trim={0cm 0cm 0cm 0cm}]{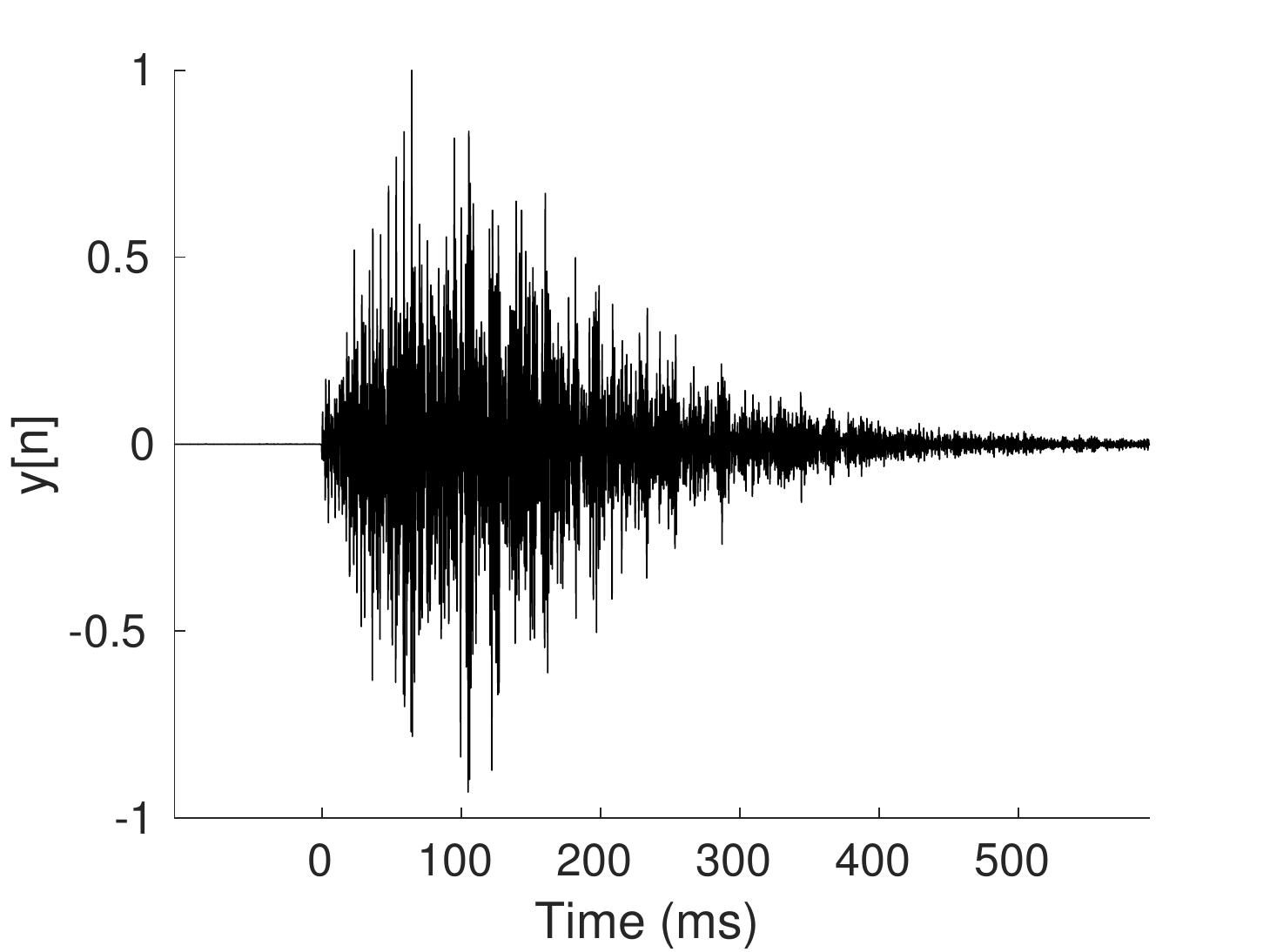}
\caption{Reverberated signal $y[n]$}
\label{fig:a_rev}
\end{subfigure}
\begin{subfigure}{0.50\textwidth}
\includegraphics[scale=0.44,trim={0cm 0cm 0cm 0cm}]{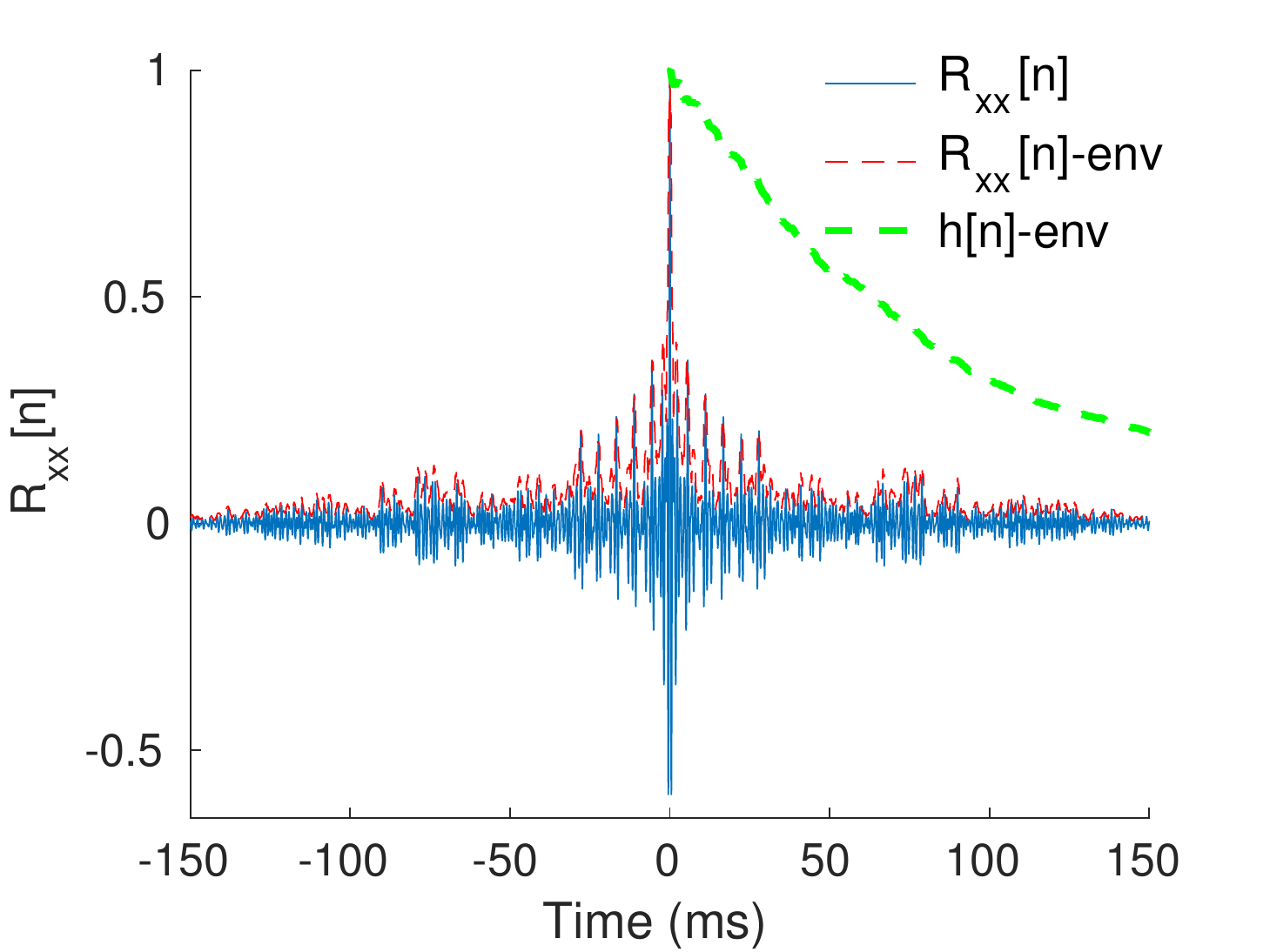}
\caption{Autocorrelation $R_{xx}[n]$}
\label{fig:a_corr}
\end{subfigure}
\begin{subfigure}{0.50\textwidth}
\includegraphics[scale=0.44,trim={0cm 0cm 0cm 0cm}]{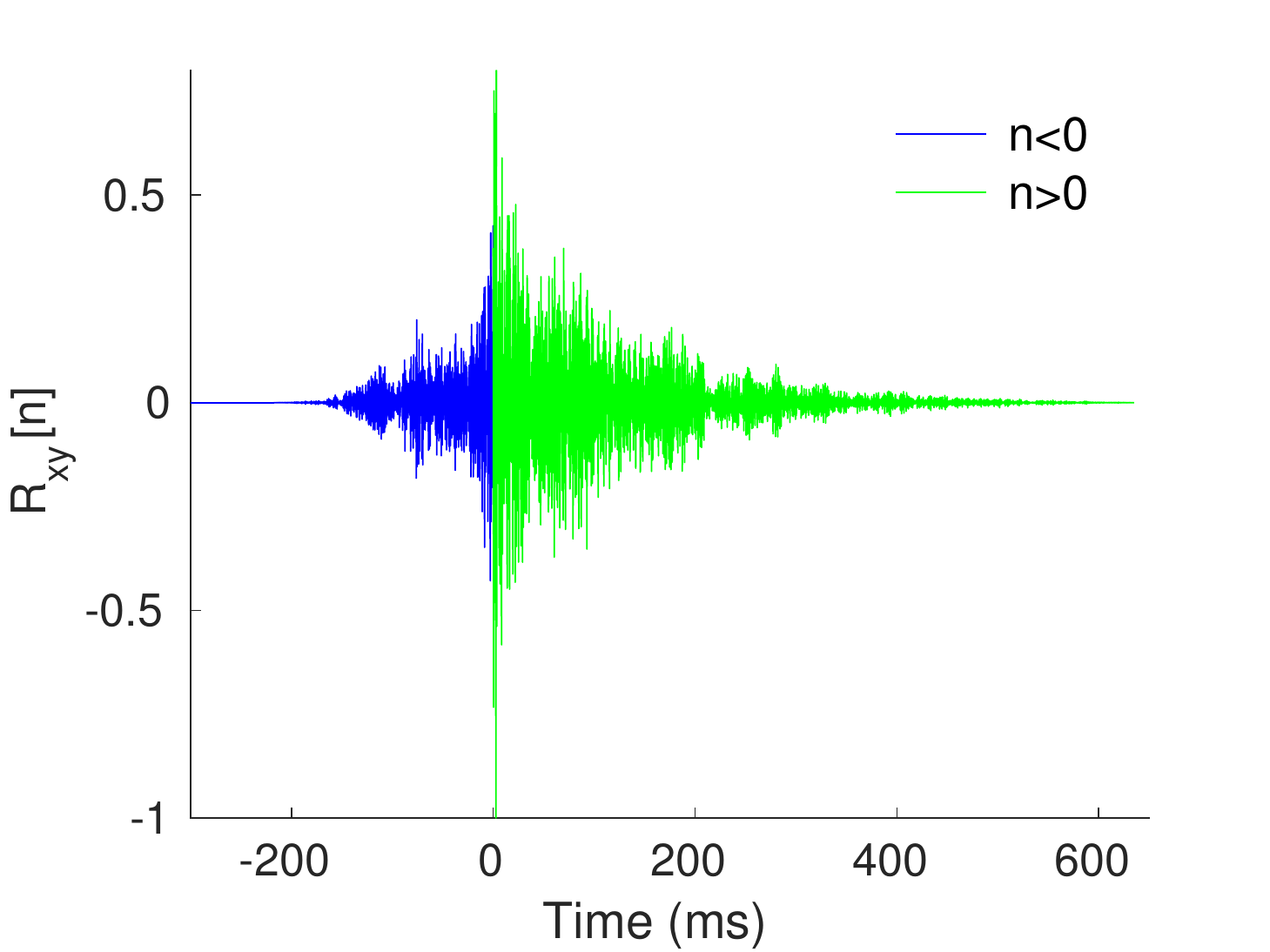}
\caption{Cross-correlation $R_{xy}[n]$}
\label{fig:a_crosscorr}
\end{subfigure}
\caption{Cross- and auto-correlation analysis for the vowel /a/. }
\label{fig:xcorr}
\end{figure}

\begin{figure}[t!]
\begin{subfigure}{0.50\textwidth}
\includegraphics[scale=0.44,trim={0cm 0cm 0cm 0cm},clip]{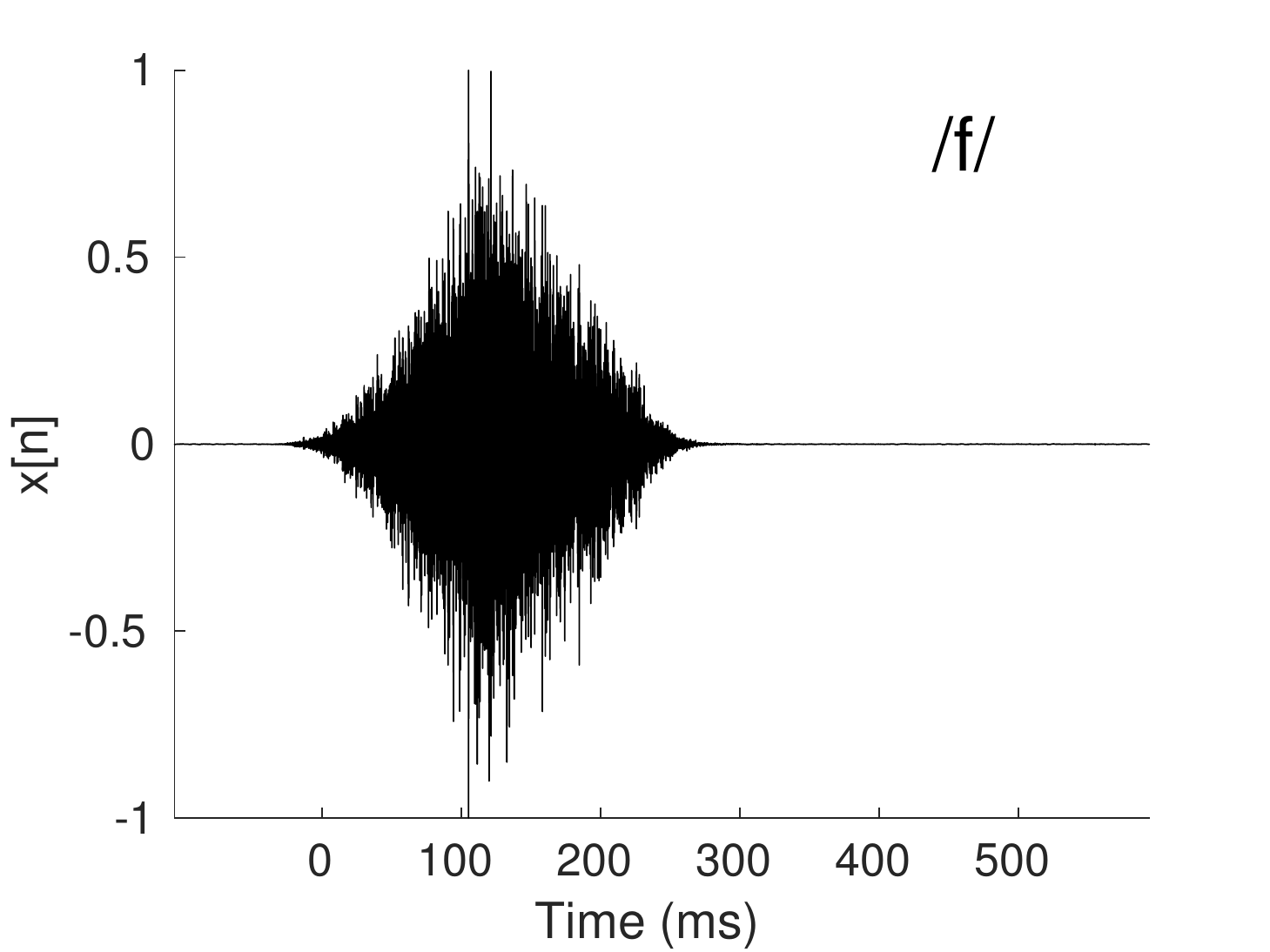}
\caption{Close-talking signal $x[n]$}
\label{fig:f}
\end{subfigure} \hspace{0.0\textwidth}
\begin{subfigure}{0.50\textwidth}
\includegraphics[scale=0.44,trim={0cm 0cm 0cm 0cm}]{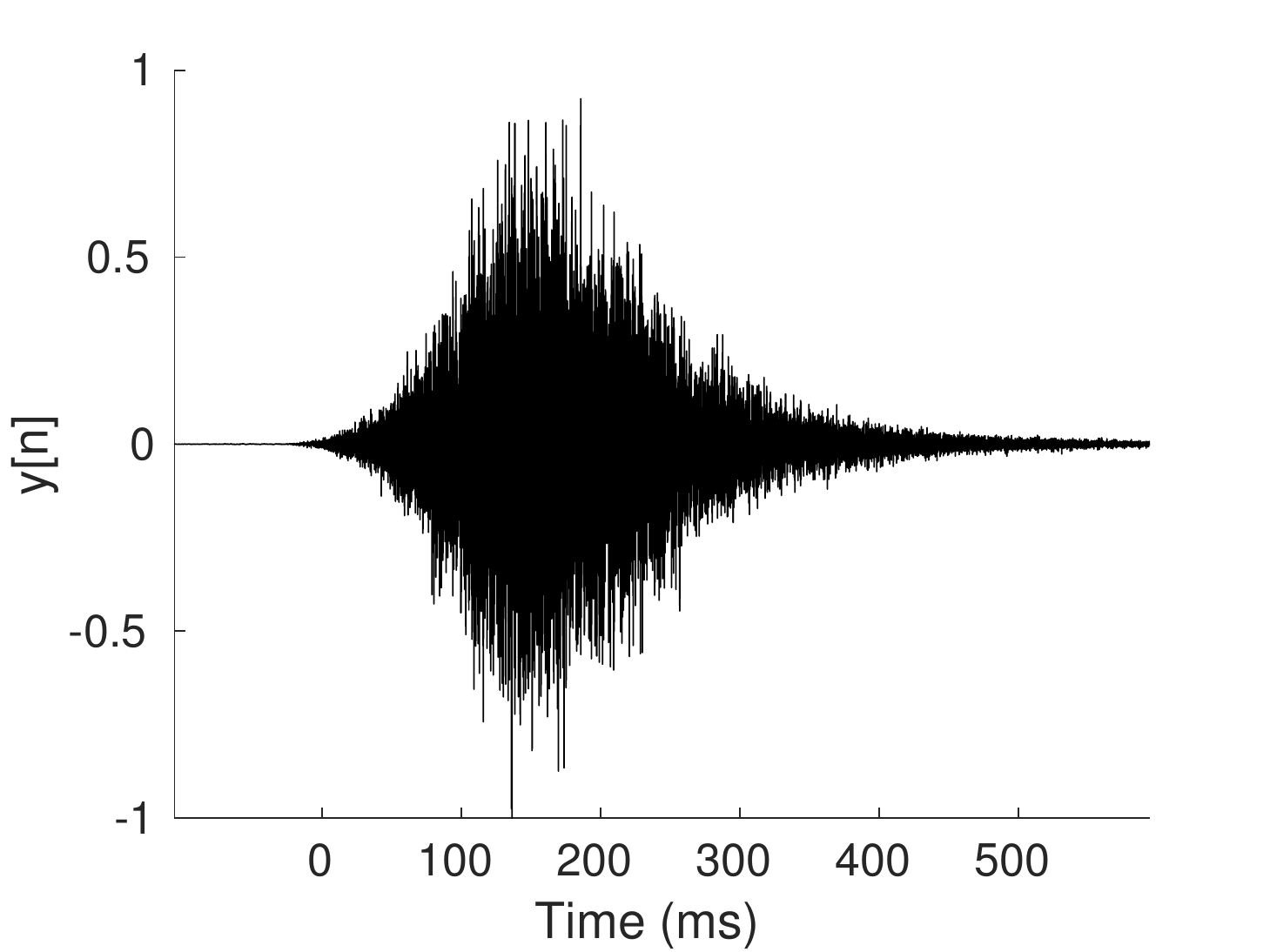}
\caption{Reverberated signal $y[n]$}
\label{fig:f_rev}
\end{subfigure}
\begin{subfigure}{0.50\textwidth}
\includegraphics[scale=0.44,trim={0cm 0cm 0cm 0cm}]{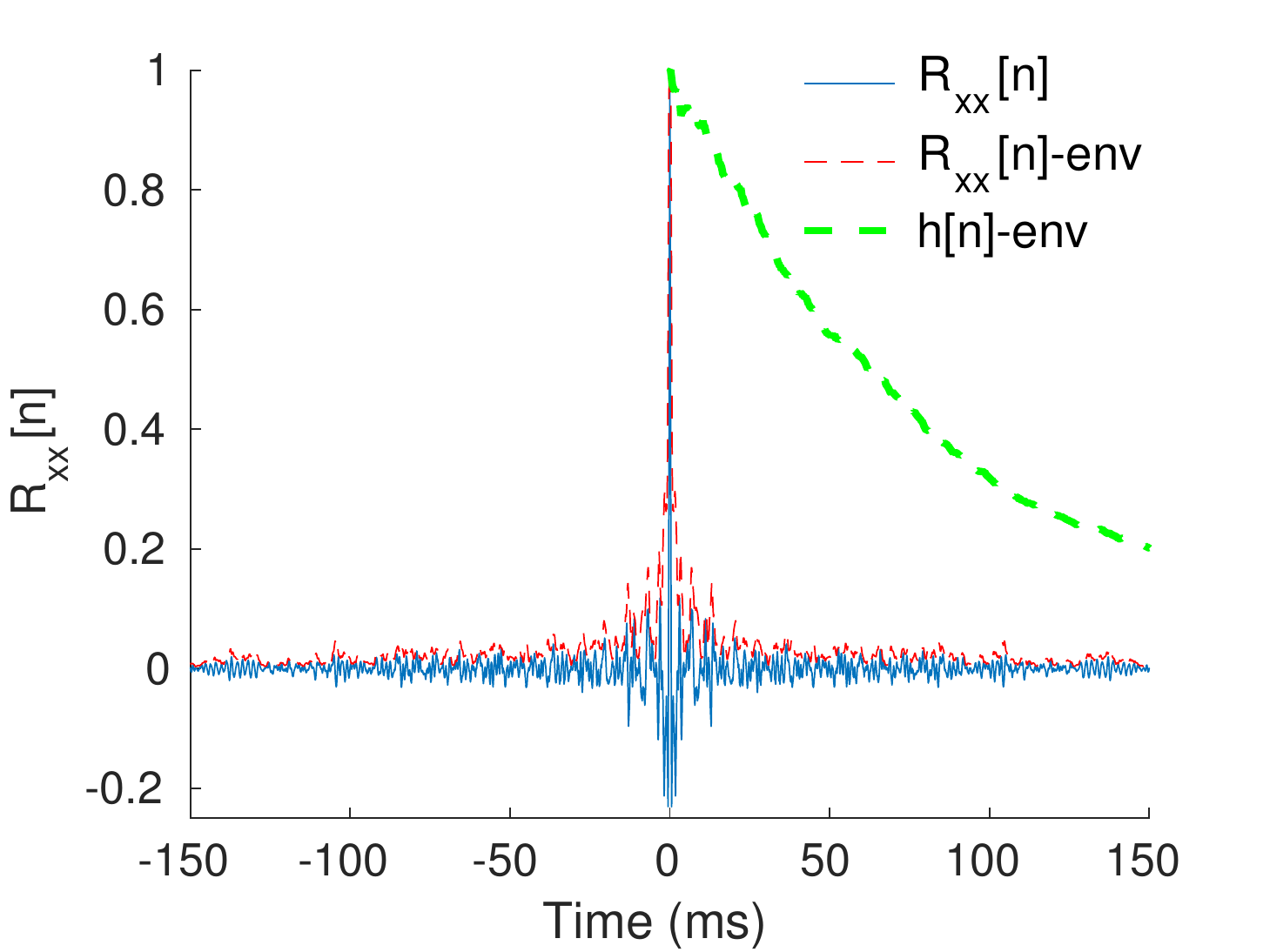}
\caption{Autocorrelation $R_{xx}[n]$}
\label{fig:f_corr}
\end{subfigure}
\begin{subfigure}{0.50\textwidth}
\includegraphics[scale=0.44,trim={0cm 0cm 0cm 0cm}]{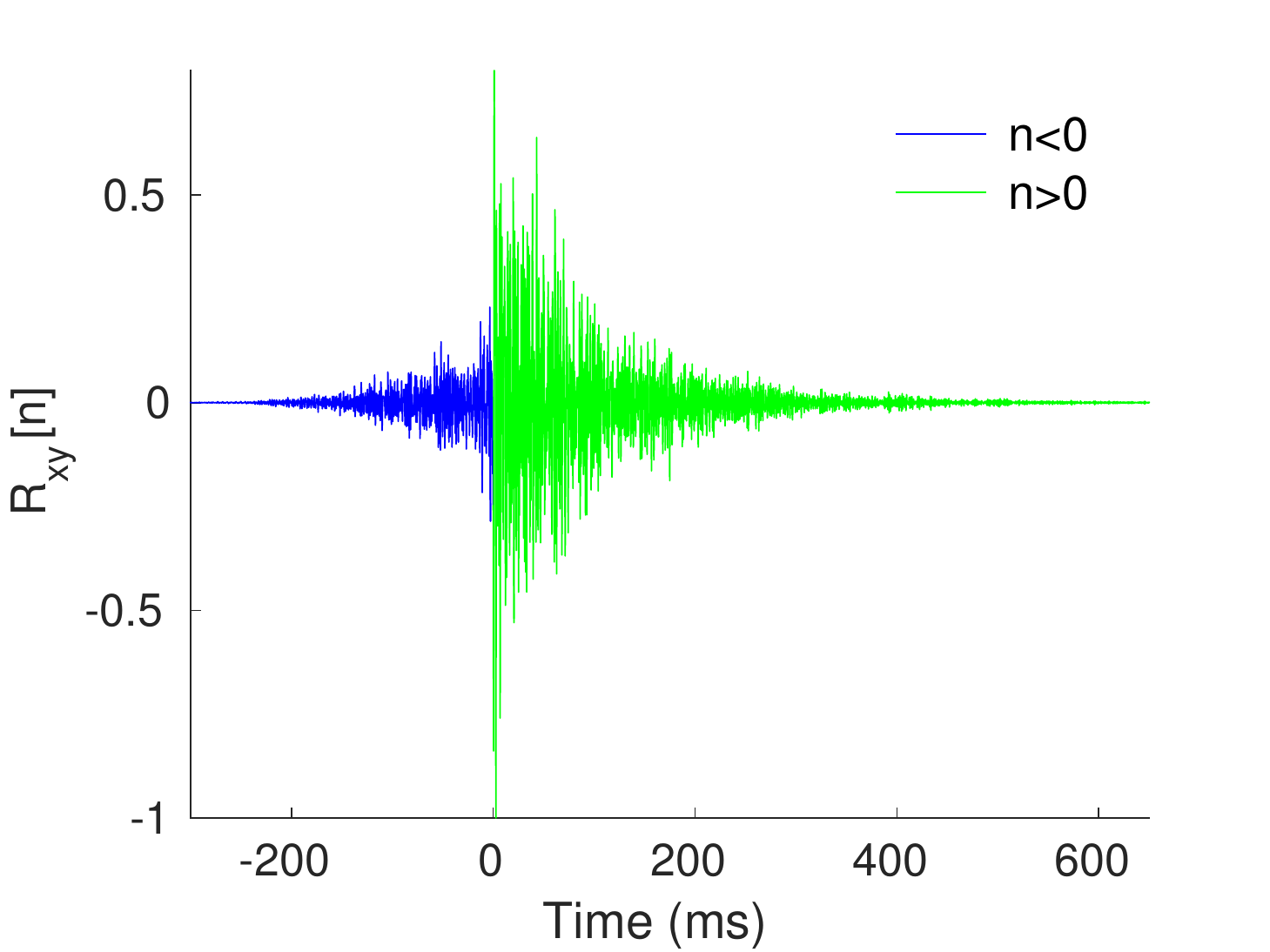}
\caption{Cross-correlation $R_{xy}[n]$}
\label{fig:f_crosscorr}
\end{subfigure}
\caption{Cross- and auto-correlation correlation analysis for the fricative /f/. }
\label{fig:xcorr2}
\end{figure}

\begin{figure*}[t!]
\centering
  \includegraphics[scale=0.61,trim={0cm 0 0cm 0},clip]{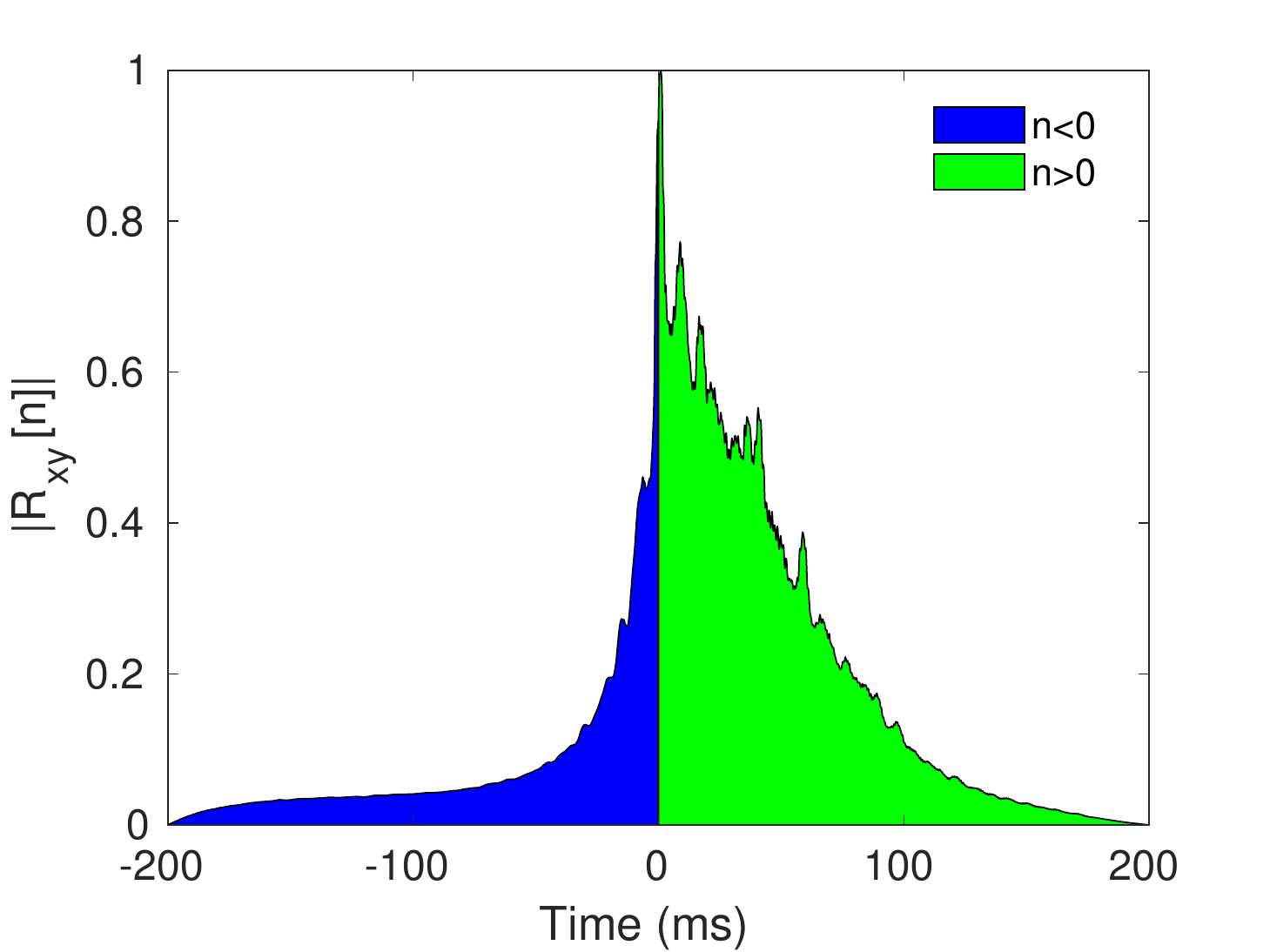}
  \caption{Envelope of the cross-correlation $R_{xy}[n]$ computed between clean and reverberated speech sentences using symmetric windows of 200 ms. The envelope $|R_{xy}[n]|$ is averaged over all the utterances of the TIMIT dataset.} 
\label{fig:ACW_avg}
\end{figure*}

%The cross-correlation $R_{xy}[n]$ thus depends on both the impulse response $h[n]$ and the autocorrelation function $R_{xx}[n]$. 
The autocorrelation $R_{xx}[n]$ varies significantly according to the particular phoneme and the signal characteristics that are considered. Fig. \ref{fig:a_corr}, for instance, shows the autocorrelation $R_{xx}[n]$ of a vowel $/a/$, while Fig. \ref{fig:f_corr} illustrates  $R_{xx}[n]$ for a fricative $/f/$. One can easily observe that different autocorrelation patterns are obtained: for the vowel sound $/a/$, $R_{xx}[n]$ is based on several peaks due to pitch and formants, while for $/f/$ a more impulse-like pattern is observed. The spread of the autocorrelation function around its center $t=0$ also depends on the specific phoneme. If we consider, for instance, the time instant where the energy of $R_{xx}[n]$ decays to 99.9\% of its initial value, the autocorrelation length is 104 ms with the vowel /a/, and about 25 ms with the fricative /f/. 
In both cases, however, the duration of the autocorrelation is significantly shorter than the IR length (see green dashed line of Fig.  \ref{fig:a_corr} and Fig. \ref{fig:f_corr}), except in the case of a very low reverberation time. 
This characteristic, together with the causality of the impulse response ($h[n]=0 \quad \forall n<0$), originates an asymmetric trend in the cross-correlation $R_{xy}[n]$, which can be clearly appreciated from both Fig. \ref{fig:a_crosscorr} and Fig. \ref{fig:f_crosscorr}. 
As shown by the latter examples, corresponding to a medium-high $T_{60}$ of 780 ms, the right side of this function is influenced by the IR decay. 
The future samples ($n>0$) are thus, on average, more redundant than previous ones ($n<0$), and this effect is amplified when reverberation increases, and in correspondence of high-energy portions of speech signals (e.g., the central part of a stressed vowel).

ACWs are therefore more appropriate than traditional symmetric ones, since they lead to a frame configuration less affected by the aforementioned forward correlation effects of reverberation. In other words,  with an asymmetric context we can feed the DSR system with information which is, on average, more complementary than that considered in a standard symmetric frame configuration, allowing the DNN to perform more robust predictions.  

As emerged from Fig. \ref{fig:xcorr}  and \ref{fig:xcorr2}, the best asymmetric context window would depend on the specific phoneme. However, this is also tightly related to the degree of distortion introduced by reverberation in the related phonetic context, which can also depend on other factors (e.g., the ratio between the energies of the direct input speech and of the reverberation component). As a matter of fact, a simple and practical solution, as outlined in the following of this work, consists in feeding the DNN with a fixed asymmetric context configuration that, on average, works reasonably well for any phonetic contexts in the input speech signal, and for different environmental conditions.

%As emerged from Fig. \ref{fig:xcorr} and  \ref{fig:xcorr2}, the best asymmetric context window would depend on the specific phoneme. Ideally, we should process different speech phones with different context windows. However, the (most practical) solution considered in this work consists in feeding the  DNN with a fixed asymmetric context configuration  that, on average, works reasonably well for all basic sounds composing the speech signal.

This approach is often adopted within standard DNN-HMM speech recognizers, where the speech signal is progressively processed using a fixed context window that might contain different sounds. To extend our cross-correlation analysis to a more realistic setting, Fig. \ref{fig:ACW_avg} shows the envelope $|R_{x,y}|$ of the cross-correlation function averaged over all the sentences of  the TIMIT dataset \cite{timit}. In particular, this result is obtained considering context windows of 200 ms and 10 ms of time shift, which resembles the typical configuration used to feed DNNs under reverberated acoustic conditions, as shown in the following of the paper. Consistently with what emerged from previous experiments, Fig. \ref{fig:ACW_avg} confirms the high redundancy introduced by reverberation on the future samples.

A similar experimental evidence can be reproduced using other methods to analyze the correlation that holds among frames inside the CW. For instance, the Pearson correlation coefficient
\cite{pearson,pearson2} could be used to highlight the redundancy inside sequences of mel-frequency cepstral coefficients (MFCC) vectors that represent reverberated speech signals. 

\subsection{Gradient Analysis}

\begin{figure}[t!]
\begin{subfigure}{0.50\textwidth}
\includegraphics[scale=0.44]{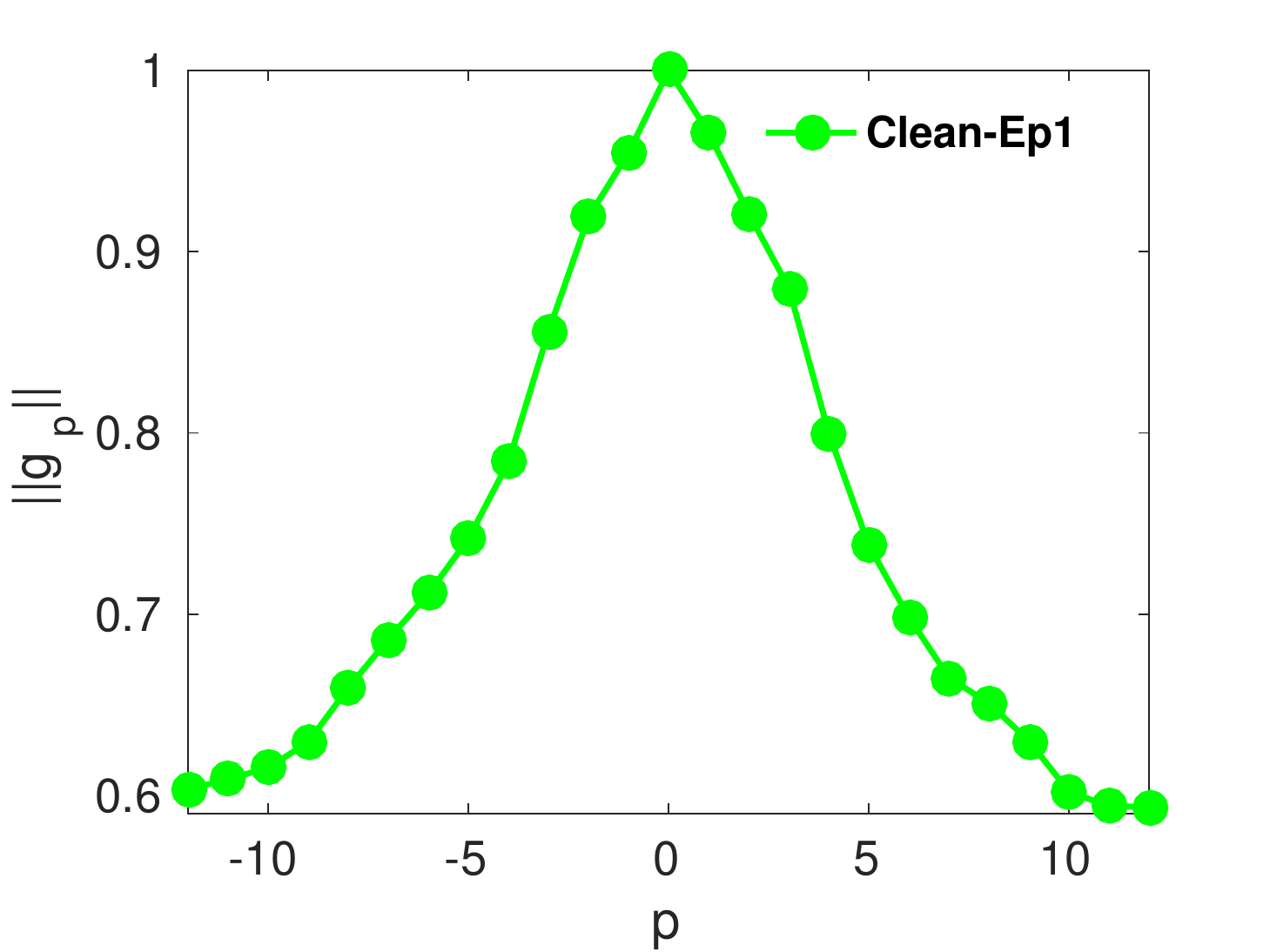}
\caption{Gradient norms at epoch 1}
\label{fig:grad_clean0}
\end{subfigure} \hspace{0.0\textwidth}
\begin{subfigure}{0.50\textwidth}
\includegraphics[scale=0.44]{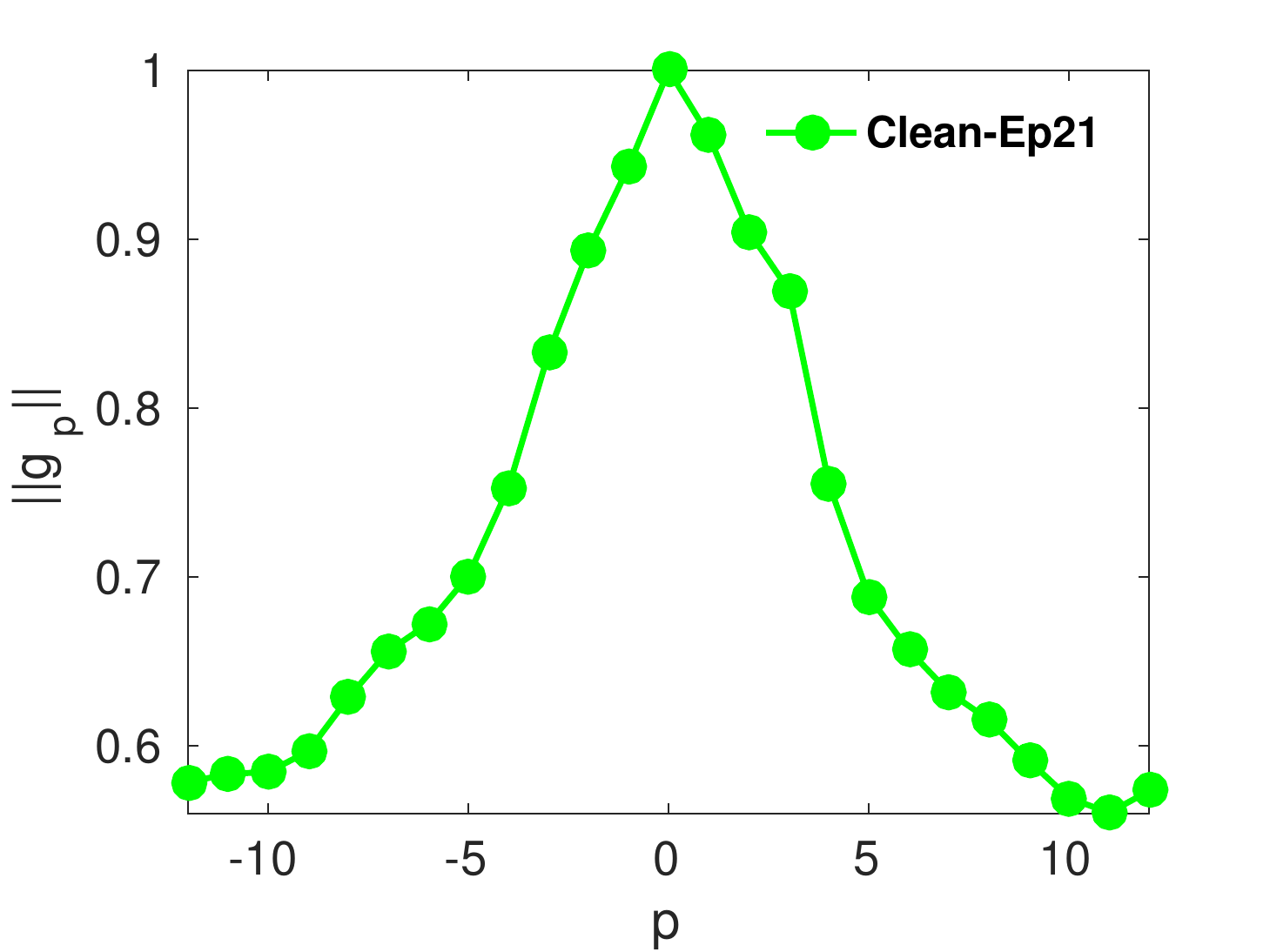}
\caption{Gradient norms at epoch 21}
\label{fig:grad_clean21}
\end{subfigure}
\caption{Gradient norm of the various frames in a close-talking scenario over various training epochs.}
\label{fig:grad_clean}
\end{figure}

\begin{figure}[t!]
\begin{subfigure}{0.50\textwidth}
\includegraphics[scale=0.44]{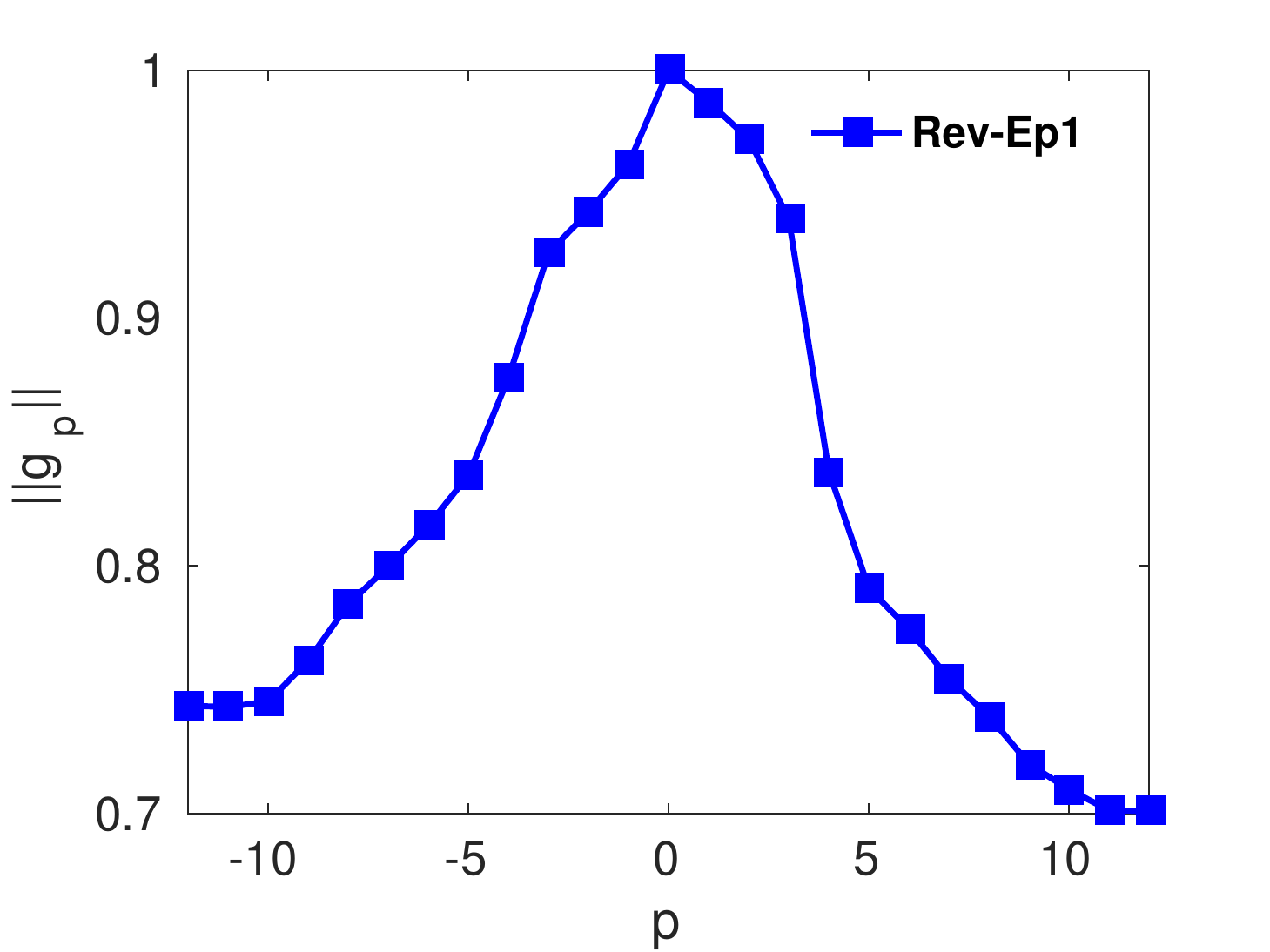}
\caption{Gradient norms at epoch 1}
\label{fig:grad_rev0}
\end{subfigure} \hspace{0.0\textwidth}
\begin{subfigure}{0.50\textwidth}
\includegraphics[scale=0.44]{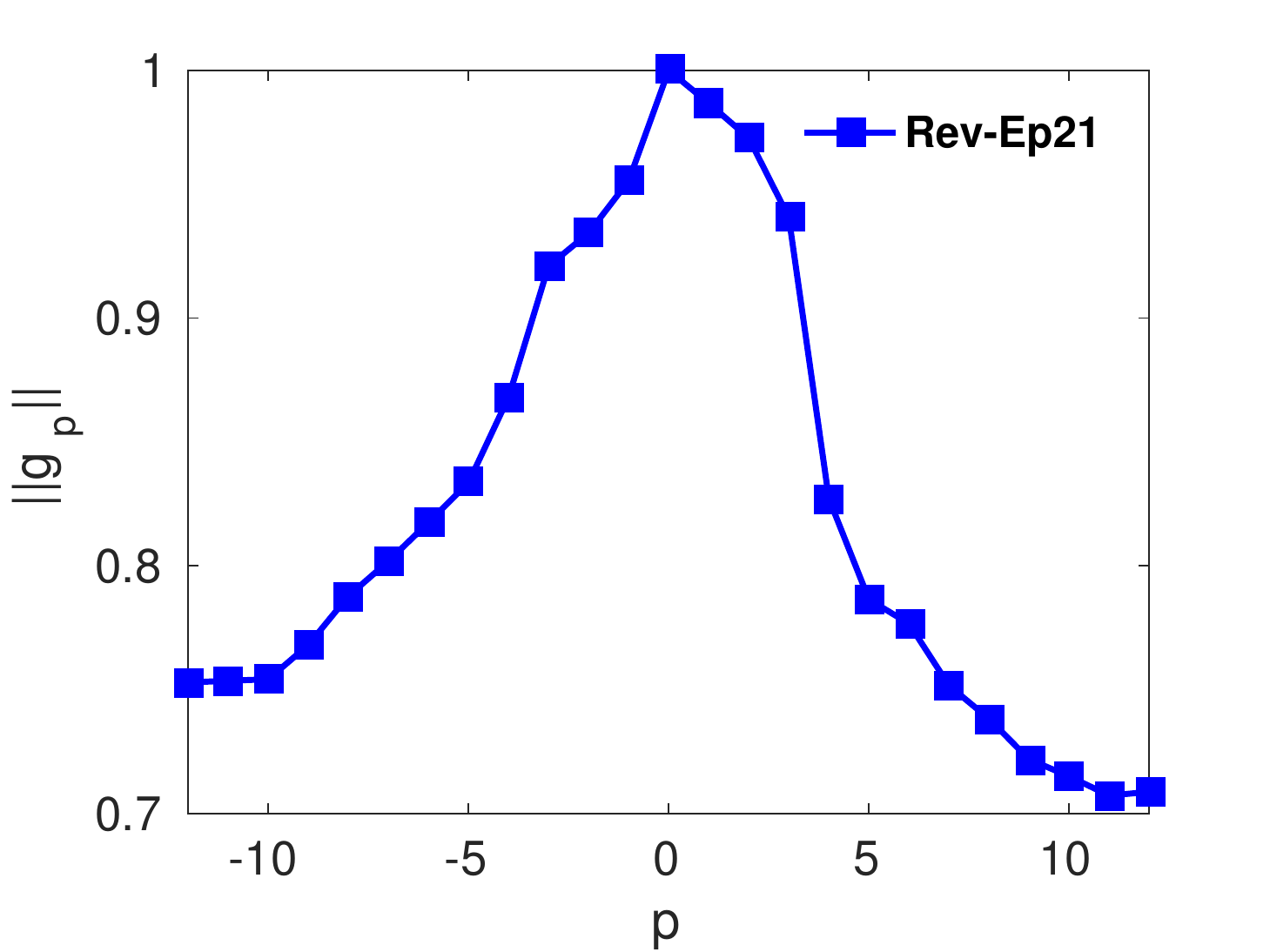}
\caption{Gradient norms at epoch 21}
\label{fig:grad_rev21}
\end{subfigure}
\caption{Gradient norm of the various frames in a distant-talking reverberated scenario over various training epochs.}
\label{fig:grad_rev}
\end{figure}

As far as DNN processing is concerned,
it is also of great interest to understand if the network is able to automatically assign different importance to the different frames of the CW. Useful insights can  be gained by analyzing the gradient norm over the various inputs of the CW, which can be defined in this way:

\begin{equation}
\norm{g_{p}}= \norm{\frac{\partial C}{\partial y_{p}}} \quad \forall p\in \{-N_{p},...,0,...,N_{f}\}
\end{equation}

where $C$ is the cost function used for DNN training and $y_{p}$ is the $p$-th feature frame embedded in the CW. In the case of cross-entropy cost, the gradient norm can be written as:

\begin{equation}
\norm{g_{p}}= \frac{1}{N_{tr}} \mathlarger{\mathlarger{\sum}}_{i=1}^{N_{tr}} \norm{\frac{\partial \sum_{k=1}^{N_{e}} \sum_{j=1}^{N_{o}}\hat{l}_{i,k,j} \cdot log \Big(P_{i,k,j}(q_{j}|\{y_{i,k,p}\}) \Big )}{\partial y_{p}}} %\quad \forall p\in \{-N_{p},...,0,...,N_{f}\}
\end{equation}

where $N_{tr}$ is the number of training mini-batches, $N_{e}$ is the number of training samples in each mini-batch, $N_{o}$ is the number of phone-states, while $\hat{l}_{i,k,j}$ and $P_{i,k,j}$ are the label and the DNN output of each training example, respectively. The DNN output $P_{i,k,j}$ depends on the context windows $CW_{i,k}$ that is written here  as $\{y_{i,k,p}\}$ to highlight its dependency on the $p$-th frame. %Note also that, to provide a more reliable estimation, the gradient norm is averaged over all the training mini-batches. 
Note also that the gradient norm is averaged over all the training mini-batches in order to provide a more reliable estimation

Fig. \ref{fig:grad_clean} and \ref{fig:grad_rev} shows $\norm{g_{p}}$ for a close-talking and a distant-talking case, respectively, which was computed by considering the first and the last training epochs. The results are derived from sequences of MFCC feature vectors, and using the DIRHA-WSJ dataset \cite{dirha_asru} with the DNN setup that will be described in Sec. \ref{sec:exp}. 

The two figures highlight that the network is able to automatically assign more importance to the current frame (p=0). In both  cases, 
%close and distant-talking conditions, 
the gradient norm $\norm{g_{p}}$ clearly decreases when progressively moving far away from the current frame. However, a symmetric behavior is observed in the close-talking case only, which means that the network has no preference for past or future information.
On the other hand, the network learns to place more importance to past ($p<0$) rather than to future frames ($p>0$) for reverberated speech. This can be readily appreciated from the asymmetric trend achieved in Fig. \ref{fig:grad_rev}, which is a further indication of the possible benefits deriving from the use of ACWs. 
%In fact, with a proper asymmetric context, we can directly feed the network with the most important frames, avoiding to overload it with useless information. Note that a  different trend is obtained with close-talking data. In the latter case, indeed, the network shows once again no clear preference for past or future information. 

Interestingly, the network learns which frames are more important since the first training epoch, as evidenced by the similar trends reported in Fig. \ref{fig:grad_clean0} and \ref{fig:grad_rev0}.
%in fact, shown that the symmetric and asymmetric trends  are attained starting from the first epoch. 
This is an important experimental evidence, which suggested us to develop the algorithm introduced in the next section, 
%In the next section, we propose an algorithm that exploits this characteristic  
to effectively optimize the hyperparameters of the ACW.

\section{Automatic context window composition} \label{sec:algo}
\begin{algorithm}[t!]
\caption{Automatic context window composition using gradient analysis.}
\label{alg:alg}
\begin{algorithmic}[1]
 \State Train a DNN  with a large symmetric context window $CW_{max}$ for one epoch.
 \State Compute the gradient norm $||g_{p}||$, $\forall p\in [-(CW_{max}-1)/2 ,(CW_{max}-1)/2]$.
 \For{$CW_{len}$ \textbf{in} \textbf{range} ($CW_{min}$,$CW_{max}$) } 
 \State $N_{p}=0$, $N_{f}=0$
 \For{$i$ \textbf{in} \textbf{range} ($CW_{len}$-1) } 
 \If{$||g_{-N_{p}-1}||$ $>$ $||g_{N_{f}+1}||$} 
 $N_{p}=N_{p}+1$
 \Else {}
 $N_{f}=N_{f}+1$
 \EndIf

 \EndFor
\State Train the DNN  with $N_{p}$ past frames and $N_{f}$ future frames.
\State Evaluate the WER performance on the dev-set.
\State Store \{$N_{p}$,$N_{f}$,WER\} for the given $CW_{len}$.
 \EndFor
\State Choose the context window with the best performance

\end{algorithmic}
\end{algorithm}

%Carefully optimizing the hyperparameters of a DNN (e.g, learning rate, number of hidden neurons, number of hidden layers, etc.) is crucial for improving the ASR performance. 
%As discussed in the previous sections, the context window is of paramount importance, and special attention should be devoted to find a proper frame configuration. 
The characteristics of the context window are of  paramount importance to improve the ASR performance. Particular attention should thus be devoted to derive a proper frame configuration, carefully optimizing (on the development set) the main features of the context window (i.e., $N_{p}$, $N_{f}$). 

A major limitation of the ACW is that it introduces two hyperparameters  (i.e., $N_{p}$ and $N_{f}$), while only one (i.e, the  total length of the context $CW_{len}$) is needed for  standard symmetric contexts.  The introduction of an additional hyperparameter has a dramatic impact on the number of combinations to test during the optimization step. A grid search over a single hyperparameter, in fact, has a linear complexity $\mathcal{O}(CW_{len})$, while the joint optimization of both $N_{p}$ and $N_{f}$ has a quadratic complexity  $\mathcal{O}(CW_{len}^2)$. For instance, if we consider a SCW, with a total length $CW_{len}$ that varies from 11 to 25 frames, only 15 DNN training experiments are necessary, against the 270 required with an exhaustive grid search. %Conversely, an exhaustive search over all the possible asymmetric configurations within this range would require 270 experiments, making such an approach very computationally demanding. 
It is thus of great interest to develop a methodology to optimize more efficiently the hyperparameters of the ACW.   

The approach proposed in this paper is based on the gradient norm analysis introduced in the previous section. The norm of the gradient over the various frames, in fact, gives quickly an idea about what frames are considered important by the network. Based on this observation, we propose the algorithm referred  to as $AutoCW$ (Alg. \ref{alg:alg}), to automatically compose the CW. The idea is to first train a DNN with a very large SCW  (e.g. 25 frames) for a single epoch. After the first epoch, the gradient norm $\norm{g_{p}}$ over the various input frames is computed. The CW is  composed by progressively embedding, at each iteration, the past or future frame that maximizes the gradient norm. The cycle is stopped when the predefined number of context frames $CW_{len}$ has been reached. A new DNN can then be trained with the CW $\{N_{p}$, $N_{f}\}$ determined by the proposed procedure, and the corresponding ASR performance is evaluated on a development data set. This operation is repeated for all the CW lengths within a predefined range ($CW_{min }\leq CW_{len} \leq CW_{max}$), and, after that, the frame configuration $\{N_{p}$, $N_{f}\}$ providing the best ASR performance is selected.

Note that this algorithm allows one to optimize the frame configuration of the asymmetric context window with a linear computational complexity, that is comparable to that required for standard symmetric windows. For each context window length $CW_{len}$, in fact, the optimal configuration $\{N_{p}$, $N_{f}\}$ is automatically inferred from the gradient norm profile, allowing one to avoid exploring the full set of context configurations.  Similarly to SCW, if we consider a context window length ranging from $CW_{min}$=11 to $CW_{max}$=25 frames, only 15 DNN training experiments are necessary to find a proper context window.

%Once analyzed the correlation at both signal and feature levels, it is interesting to inspect the DNN weights. With this regard, the frame importance $I_{p}$, defined in Eq. \ref{eq:imp}, is reported in  Fig. \ref{fig:inside_dnn}. The frame importance $I_{p}$, which is normalized by its maximum value for representational convenience, is computed in both clean (Fig. \ref{fig:inside_dnn_clean}) and reverberated (Fig. \ref{fig:inside_rev}) conditions for MFCC features. %The index $p=0$ represents the current frame, while negative and positive values of p refer to past and future frames, respectively. 

% \begin{figure*}[t!]
% \centering
%   \includegraphics[scale=0.65]{inside_DNN.pdf}
%   \caption{} 
% \label{fig:inside_dnn}
% \end{figure*}

%To better understand the effect of reverberation inside the DNN, it is helpful to analyze its weights. In particular, we examined the weights connecting the input frames with the neurons of the first hidden layer. 

%while $w_{k,i,j}$ is the weights connecting the i-th features of the k-th frame with the j-th neurons of the first hidden layer.

%As outlined in Sec. \ref{sec:prior}, this frame configurations is also more suitable than other windows  or real-time/low-latency ASR applications, since the delay introduced by waiting future frames is minimized.

\section{Experimental Setup} \label{sec:ct}
%{\color{blue} INTRO DA RIVEDERE}

The experimental framework developed in this work is based on the use of 
%The speech recognition framework considered in this work relies on evaluation conducted with 
both WSJ-5k and LibriSpeech tasks. To provide an accurate analysis of the proposed approach, the experiments are performed under three different acoustic conditions of increasing complexity: close-talking (\textit{Clean}), distant-talking with reverberation (\textit{Rev}), and distant-talking with both noise and reverberation (\textit{Rev\&Noise}).
The corpora used for each experimental condition are summarized in Table \ref{tab:dataset} and described in the two following sections. The adopted ASR setup will be described in Sec. \ref{sec:ASR}.

\begin{table}[t!]
\centering
\tabcolsep=0.25cm
    \begin{tabular}{| l | l | l | c | c | c | c |}
\hline
Acoustic Condition & Training & Test \\ \hline
Close-talking (Clean) & WSJ-clean& DIRHA-WSJ-clean \\ \hline
Close-talking (Clean) & LibriSpeech & LibriSpeech  \\ \hline
Distant-talking (Rev) & WSJ-rev& DIRHA-WSJ-rev \\ \hline
Distant-talking (Rev) & Rev-LibriSpeech & Rev-LibriSpeech \\ \hline
Distant-talking (Rev\&Noise) & WSJ-rev& DIRHA-WSJ-rev\&noise \\ \hline
\end{tabular}
\caption{List of the experimental tasks considered in this work with the related training and test datasets.}
\label{tab:dataset}
\end{table}

\subsection{Close-talking experiments}
\label{sec:ct_data}
For close-talking experiments, we consider the standard WSJ dataset (i.e., WSJ-clean) for training, and the close-talking portion of the DIRHA English WSJ Dataset (i.e., DIRHA-WSJ-clean) for test purposes. The latter dataset was acquired during the DIRHA project in a recording studio of FBK, using professional equipment to obtain high-quality speech material \cite{dirha_asru}.
In this work, we used a subset of the corpus, consisting of 409 WSJ sentences (with the same text used for the CHiME \cite{chime3} challenge) uttered by six US speakers (three males and three  females). 

To evaluate the proposed method on a larger scale ASR task, additional experiments were performed with the LibriSpeech dataset \cite{librispeech}, that is based on speech material derived from read audio-books. In particular, we used a training subset consisting of 460 hours of speech uttered by 1172 speakers. %carefully segmented and aligned.
% attenzione il numero degli speaker in librispeech non puo' essere preso dal file spk2gen perche ci sono dei doppioni. Il numbero vero degli speaker e' quello riportato nella versione corrente dove il numero degli speakers coincide con quanto riportato nel paper di riferimento di librispeech.

%\begin{table}[t!]
%\centering
%\tabcolsep=0.30cm
%    \begin{tabular}{| l | c | c | c | c | c | c |}
%\hline
%Dataset & Use & Rev. & Noise & Data Type  \\ \hline
%WSJ-clean & Train & no & no & Close-talking  \\ \hline
%DIRHA-WSJ-clean & Test & no & no & Close-talking  \\ \hline
%WSJ-rev & Train & yes & no & Distant-talking (sim)\\ \hline
%DIRHA-WSJ-rev & Test & yes & no & Distant-talking (sim)\\ \hline
%DIRHA-WSJ-rev\&noise & Test & yes & yes & Distant-talking (real)\\ \hline
%\end{tabular}
%\caption{List of the main corpora adopted in this work.}
%\label{tab:dataset}
%\end{table}

\subsection{Distant-talking experiments}
\label{sec:dt_data}
The reference environment for several experiments conducted in this study is the living-room of a real apartment (available under the DIRHA project) with a reverberation time $T_{60}$ of about $780$ ms. The living-room was equipped with a microphone network composed of 40 microphones.
An IR measurement session, exploring a large number of positions and orientations of the sound source, was conducted in the aforementioned targeted environment with the purpose of generating realistic simulated data. More information on the adopted IR estimation procedure can be found in \cite{rav_is16,ravanelli}. 

%{\color{blue} DA RIVEDERE USO DEI TEMPI QUI AL PRESENTE, ALTROVE IN GENERALE AL PASSATO}

A set of experiments is carried out to study distant-talking conditions where only reverberation acts as a source of disturbance (Rev).  In this case, training is performed using a contaminated dataset (i.e., WSJ-rev), which is generated by convolving the original WSJ-clean data set with a set of three IRs chosen from the aforementioned collection. 
The corresponding test data set, i.e. DIRHA-WSJ-rev, is based on a contaminated version of  DIRHA-WSJ-clean. In  order  to  simulate several speaker positions and orientations, a set of 36 IRs (different from those used for training) is used for the latter dataset. %Such simulated test dataset, will be referred to as DIRHA-WSJ-rev.

To explore more challenging conditions characterized by both noise and reverberation (Rev\&Noise), real recordings have also been performed.
The real recordings,  referred to as DIRHA-WSJ-rev\&noise, are part of the recently-released DIRHA English WSJ corpus \cite{dirha_asru} and are composed of 409 WSJ sentences (with the same texts used to record DIRHA-WSJ-clean) uttered by six US speakers. Each subject reads a set of WSJ sentences from a tablet, standing still or sitting on a chair. Every 11-12 sentences, he/she was asked to move to a new position and take another orientation. %For each speaker, the read material corresponds to the same list of texts used to record the clean speech dataset (DIRHA-WSJ-clean). 
Different typologies of non-stationary domestic noises affect the signals (e.g., vacuum cleaner, microwave noise, interfering speakers talking in other rooms, kitchen tools, open window noises,etc.), resulting in an average SNR of about 10 dB (for more details see \cite{dirha_asru}\footnote{This dataset is distributed by the Linguistic Data Consortium (LDC).}).

To test our approach in different contexts, other contaminated versions of the training and test data are generated with different IRs (either measured in other real environments, or computed with the image method \cite{image}), as discussed in Sec. \ref{sec:mis} and Sec. \ref{sec:t60}.

Other experiments are performed with a reverberated version of the LibriSpeech dataset \cite{librispeech}. %that is based on speech material derived from read audio-books. %carefully segmented and aligned.  
The original close-talking sentences are convolved with 2145 IRs, that are measured in various positions and with different microphone configuration of the aforementioned living-room. The two test sets (here denoted  as \textit{Test1} and \textit{Test2}), are composed of 2620 sentences uttered by 40 speakers, and 2939 sentences uttered by 33 speakers, respectively. The test sentences are convolved with about 2000 IRs, corresponding to speaker positions and microphones different from those used for training. Note that the test data of the Librispeech corpus are originally clustered so that lower-WER speakers are gathered into $\textit{Test1}$, while the others are in $\textit{Test2}$. 

\subsection{DNN and ASR setup}
\label{sec:ASR}
In this work, we use a context-dependent DNN-HMM speech recognizer, where every unit is modeled by a three state left-to-right HMM, and the tied-state observation probabilities are estimated through a DNN. 

%{\color{blue} AGGIUNGEREI QUALCOSA QUI A PROPOSITO DI DNN, COM'É, COME VENGONO STIMATI I PARAMETRI...}

Feature extraction is based on splitting the signal into frames of 25 ms with an overlap of 10 ms.  The experimental activity is conducted considering different acoustic features, i.e., 39 MFCCs (13 static+$\Delta$+$\Delta\Delta$), 40 log-mel filter-bank features (FBANKS), as well as 40 fMLLR features (extracted as reported in the s5 recipe of Kaldi \cite{kaldi}). Features of consecutive frames are gathered into both symmetric and asymmetric observation windows. 
%Mean and variance normalization of the feature space is applied before feeding the DNN. 
As for MFCCs, it is worth mentioning that one could conduct this study without using derivatives. In the latter case, the experimental results would be quite similar at qualitative level. In other words, we would obtain a trend that reflects what reported in the following section, though with a more prominent relative decrease of performance when adopting a non-optimal context length settings, because of 
%the different 
a less effective
way the contextual information is exploited. For this reason, here we prefer to report results related to the use of the first and second order derivatives.

WSJ experiments are based on DNNs composed of six sigmoid-based hidden layers of 2048 neurons, that are trained with the Kaldi toolkit \cite{kaldi} (Karel's recipe). %while the output is based on a softmax classifier. 
Weights are initialized with the standard Glorot initialization \cite{xavier}, while biases are initialized to zero.  
Training is performed with Stochastic Gradient Descend (SGD)
that optimizes the cross-entropy loss function. The training evolution is monitored using a small validation set (10\% of the training data) that is randomly extracted from the training corpus. The performance on the validation set is monitored after each epoch to perform learning rate annealing  as well as for checking the stopping condition.   
In particular, the initial learning rate is kept fixed at 0.008 as long as the increment of the frame accuracy on the validation is higher than 0.5$\%$. For the following epochs, the learning rate is halved until the increment of frame accuracy is less than the stopping threshold of 0.1$\%$. The labels for DNN training are derived from an alignment on the tied states, which is performed with a previously-trained HMM-GMM acoustic model \cite{kaldi}. %with the same input features and adopting the same corpus. 
For Convolutional Neural Network (CNNs) experiments, we replace the first two fully-connected layers of the above-mentioned DNN with two convolutional layers based on 128 and 256 filters, respectively.

Librispeech experiments rely on the standard $nnet2$ implementation of Kaldi, which employs a generalized maxout network (p-norm). In particular, our experiments are based on a four hidden layer p-norm architecture trained for 10 epochs with minibatches of size 128. The initial learning rate is set to 0.01, while the final one is 0.001. See \cite{pnorm} and the kaldi recipe in $Librispeech/s5$ for more details.

%Unless differently specified, the hyperparameters of the CW (i.e., $CW_{len}$ for SCW, and $N_{p}$, $N_{f}$ for ACW) have been optimized using a grid search approach. 

\section{ASR Results}
\label{sec:exp}
In the following, we report the experimental results obtained on the addressed ASR tasks.  In Sec. \ref{sec:acw_exp}, a comparison between SCWs and ACWs is conducted considering different context configurations, input features as well as DNN architectures. In Sec. \ref{sec:mis}, we test the performance of the proposed $AutoCW$ algorithm with different recognition tasks and real acoustic environments,  while in Sec. \ref{sec:t60} we extend the speech recognition validation by simulating different reverberation times.

% \begin{figure}[t!]
% \begin{subfigure}{0.50\textwidth}
% \includegraphics[scale=0.44]{cepstral_clean.pdf}
% \caption{Close-talking scenario (Clean)}
% \label{fig:clean_cep}
% \end{subfigure} \hspace{0.0\textwidth}
% \begin{subfigure}{0.50\textwidth}
% \includegraphics[scale=0.44]{cepstral_rev.pdf}
% \caption{Distant-talking scenario (Rev)}
% \label{fig:rev_cep}
% \end{subfigure}
% \caption{Cepstral Distance (d) across the various frames composing the context windows}
% \label{fig:fea_an}
% \end{figure}

%\subsection{ASR performance analysis}
\subsection{Reverberant speech recognition with asymmetric context windows}
\label{sec:acw_exp}
\begin{figure}[t!]
\begin{subfigure}{0.50\textwidth}
\includegraphics[scale=0.44]{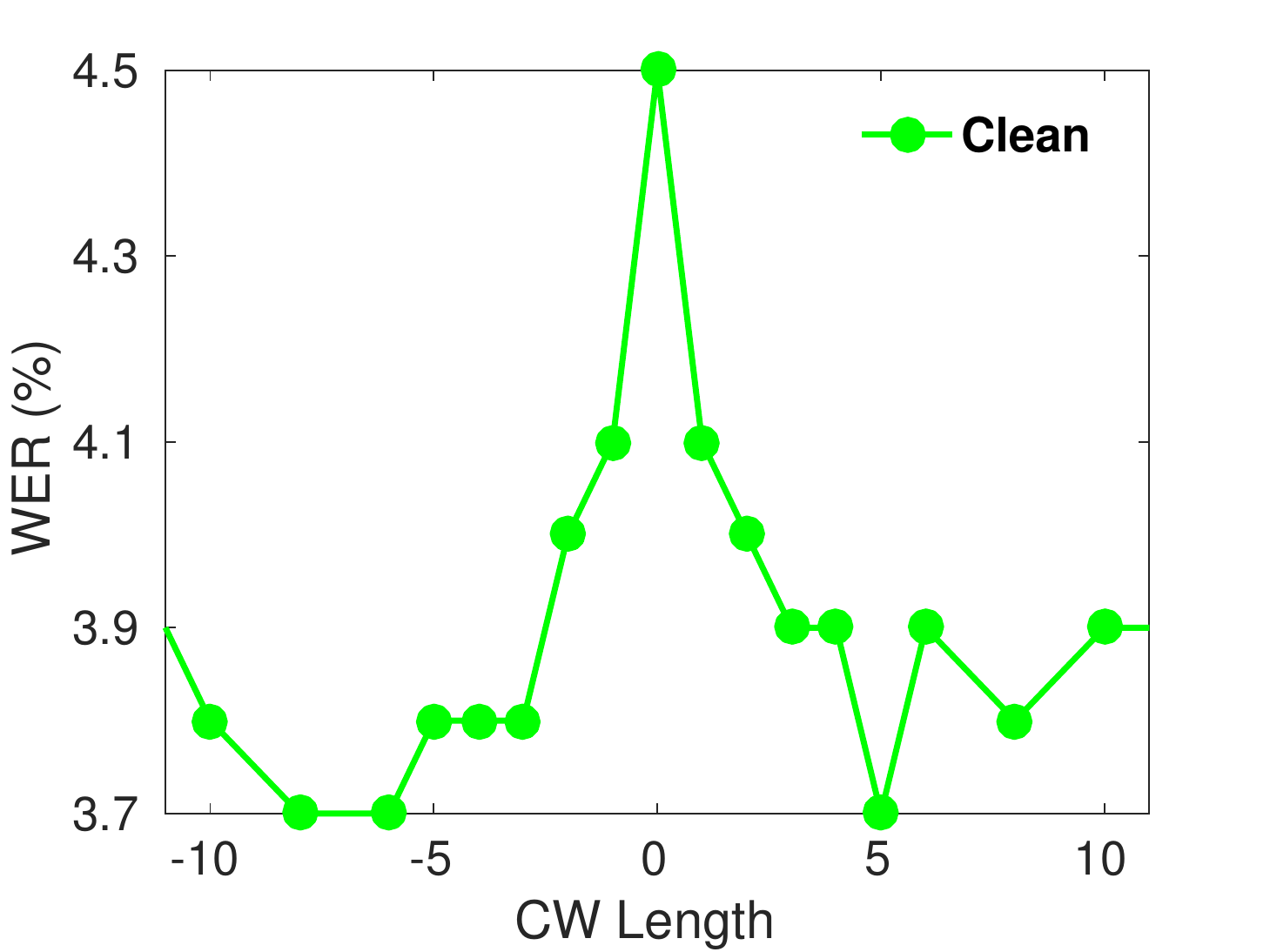}
\caption{Close-talking scenario (Clean)}
\label{fig:past_future_clean}
\end{subfigure} \hspace{0.0\textwidth}
\begin{subfigure}{0.50\textwidth}
\includegraphics[scale=0.44]{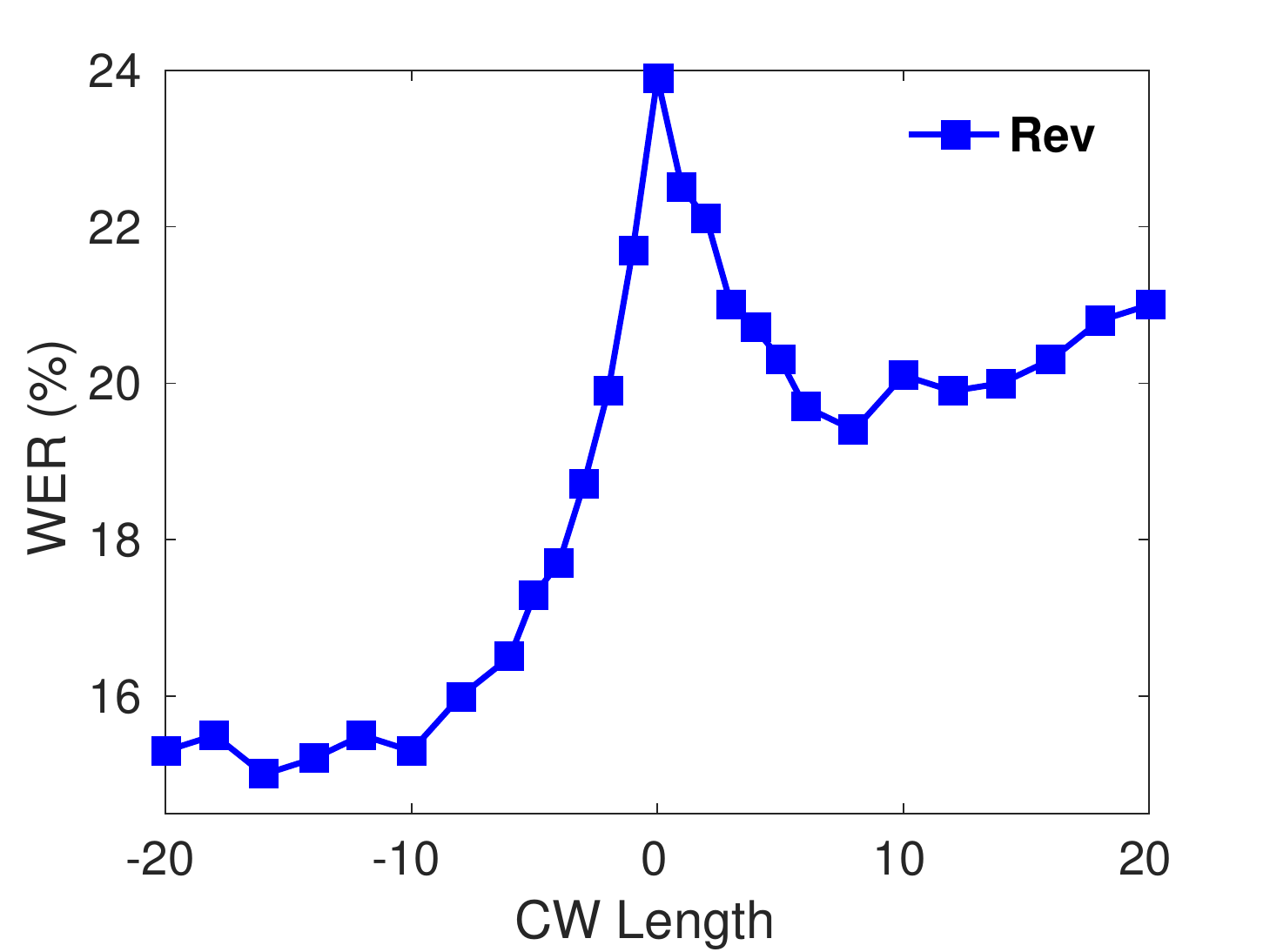}
\caption{Distant-talking scenario (Rev)}
\label{fig:past_future_rev}
\end{subfigure}
\caption{WER(\%) obtained with DNN context windows that progressively integrate only past or future frames (using fMLLR features). Results refer to the use of DIRHA-WSJ-clean (a), and DIRHA-WSJ-rev (b), tasks.}
\label{fig:past_future}
\end{figure}

From the preliminary study on ACWs, carried out in the previous sections, we found that the training of distant-talking DNNs tends to naturally attribute more importance to past rather than future frames. In this section, we take a step forward by verifying whether this fact is also observed in terms of recognition performance.
With this purpose, Fig. \ref{fig:past_future} shows the Word Error Rate (WER) results obtained in close-talking (Clean) and reverberant (Rev) conditions, when using fully asymmetric (i.e., single side) context windows of different lengths. Negative x-axis refers to the progressive integration of past frames only ($\rho_{cw}$=100\%), while positive x-axis refers to future frames ($\rho_{cw}$=0\%). In this set of experiments, fMLLR features were used as input to the DNN, for both DIRHA-WSJ-clean and DIRHA-WSJ-rev tasks. Note that similar trends have been obtained with both MFCCs and FBANK features.

Results highlight that a rather symmetric behavior is attained in the close-talking case (Fig. \ref{fig:past_future_clean}), reiterating that in such contexts past and future information provides a similar contribution to improve the system performance. %A standard symmetric context window observing a total of 11 frames seems appropriate in such a noiseless scenario. 
Differently, the role of past information is significantly more important in the distant-talking case, since a faster decrease of the WER(\%) is observed when past frames are progressively concatenated (Fig. \ref{fig:past_future_rev}). 
This result is in line with the findings emerged in the previous sections, and it confirms that an ACW is more suitable than a traditional symmetric one when reverberation arises.
%Although this results further confirm the experimental evidence emerged in the previous section, it could be of inter

\begin{figure}[t!]
\begin{subfigure}{0.50\textwidth}
\includegraphics[scale=0.44]{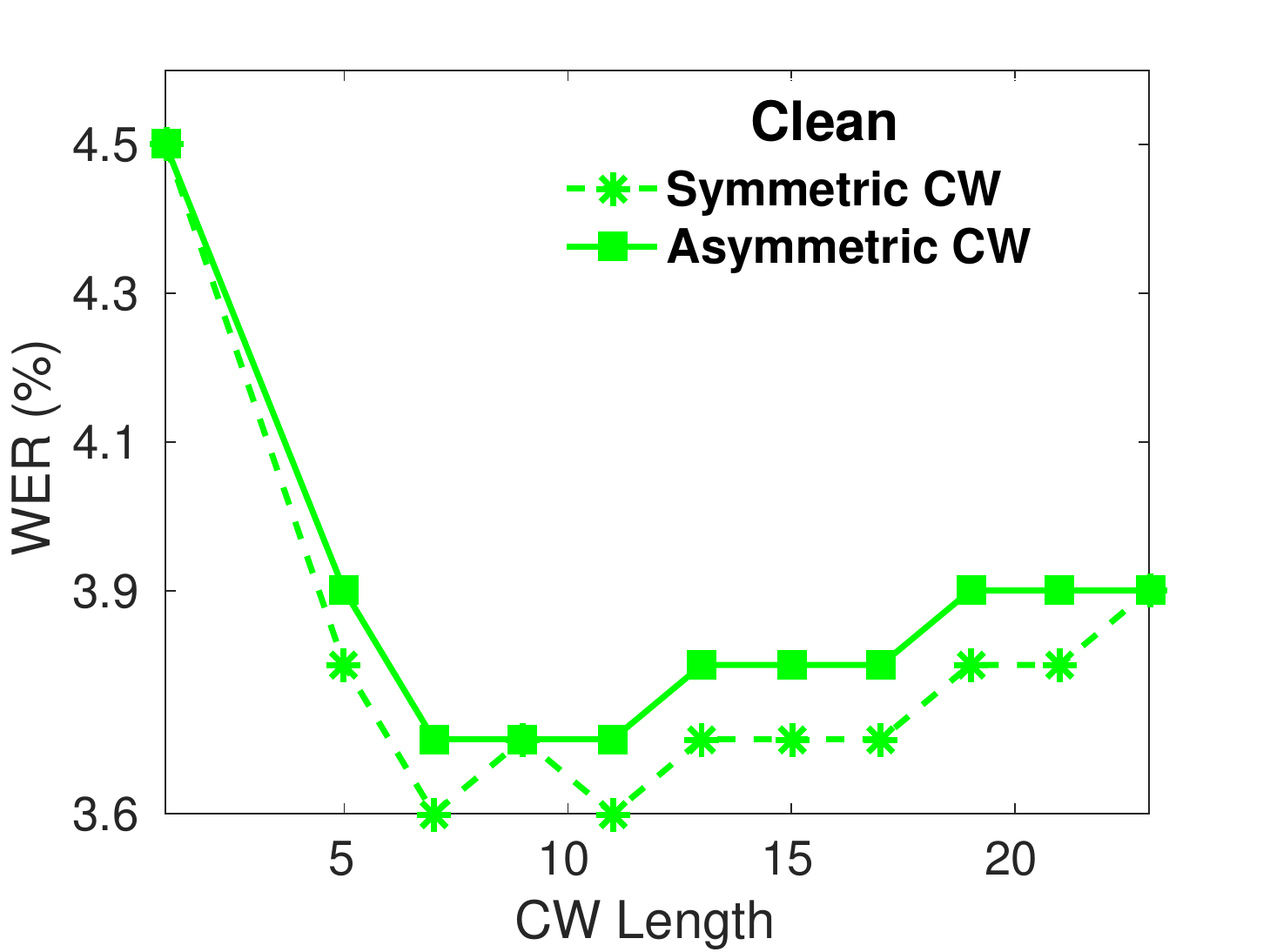}
\caption{Close-talking scenario}
\label{fig:acw_cl}
\end{subfigure} \hspace{0.0\textwidth}
\begin{subfigure}{0.50\textwidth}
\includegraphics[scale=0.44]{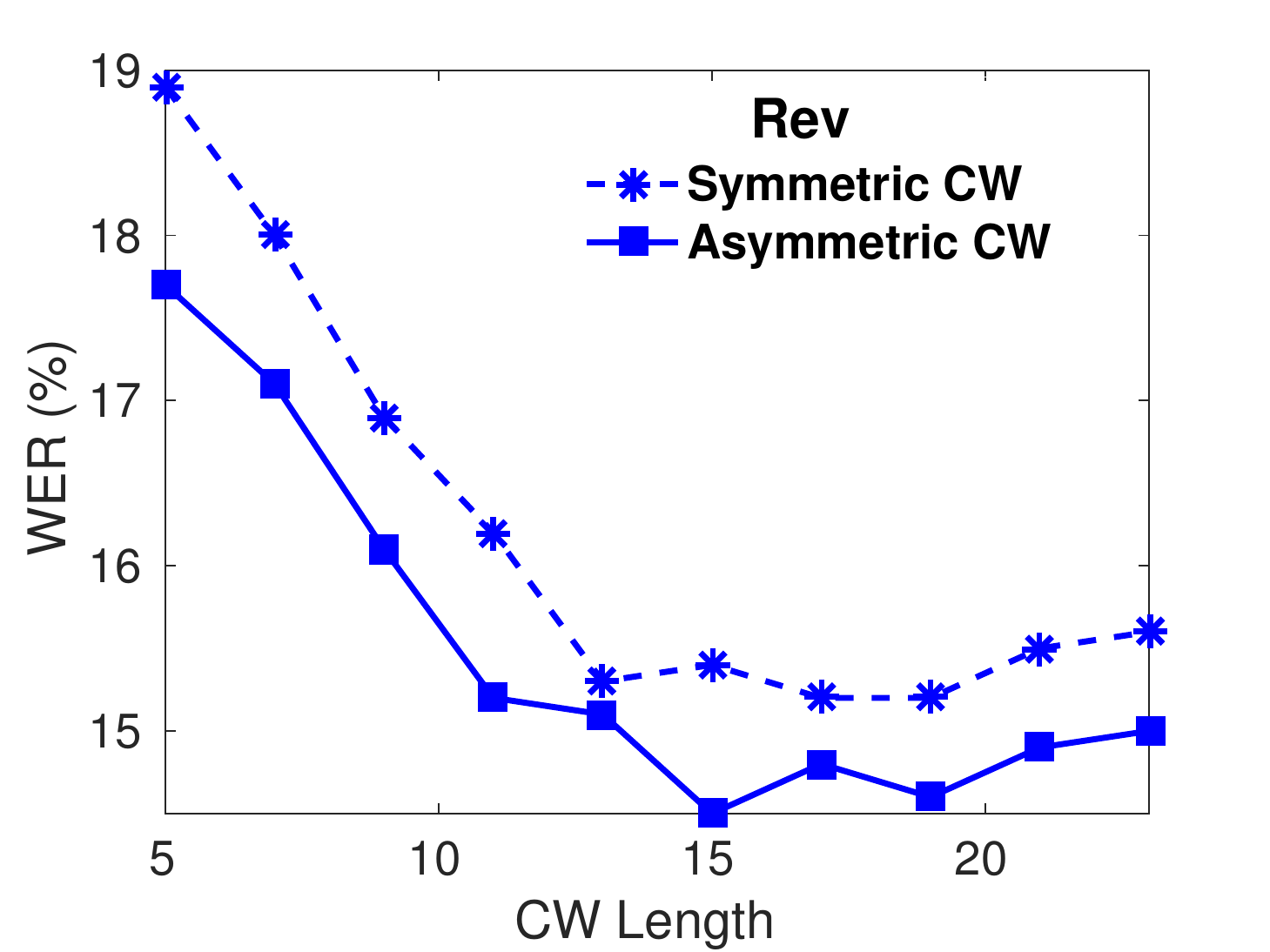}
\caption{Reverberated Scenario}
\label{fig:acw_rev}
\end{subfigure}
\caption{Comparison between SCW and ACW under both a close-talking and distant-talking reverberated conditions (using fMLLR features). Results refer to the used of DIRHA-WSJ-clean (a), and DIRHA-WSJ-rev (b) tasks.}
\label{fig:acw_scw}
\end{figure}

\begin{figure}[t!]
\begin{subfigure}{0.50\textwidth}
\includegraphics[scale=0.44]{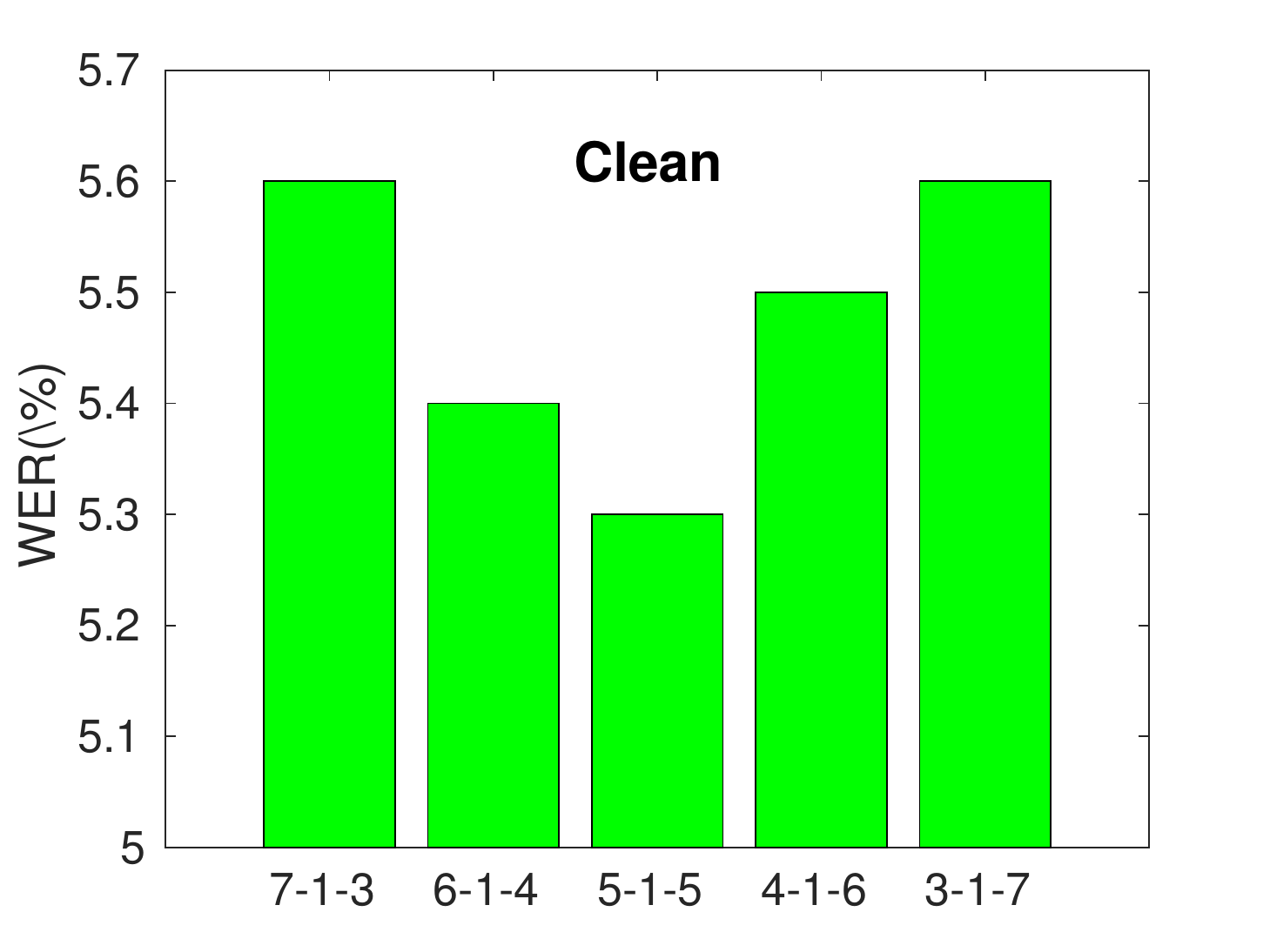}
\caption{Close-talking scenario}
\label{fig:libri_cl}
\end{subfigure} \hspace{0.0\textwidth}
\begin{subfigure}{0.50\textwidth}
\includegraphics[scale=0.44]{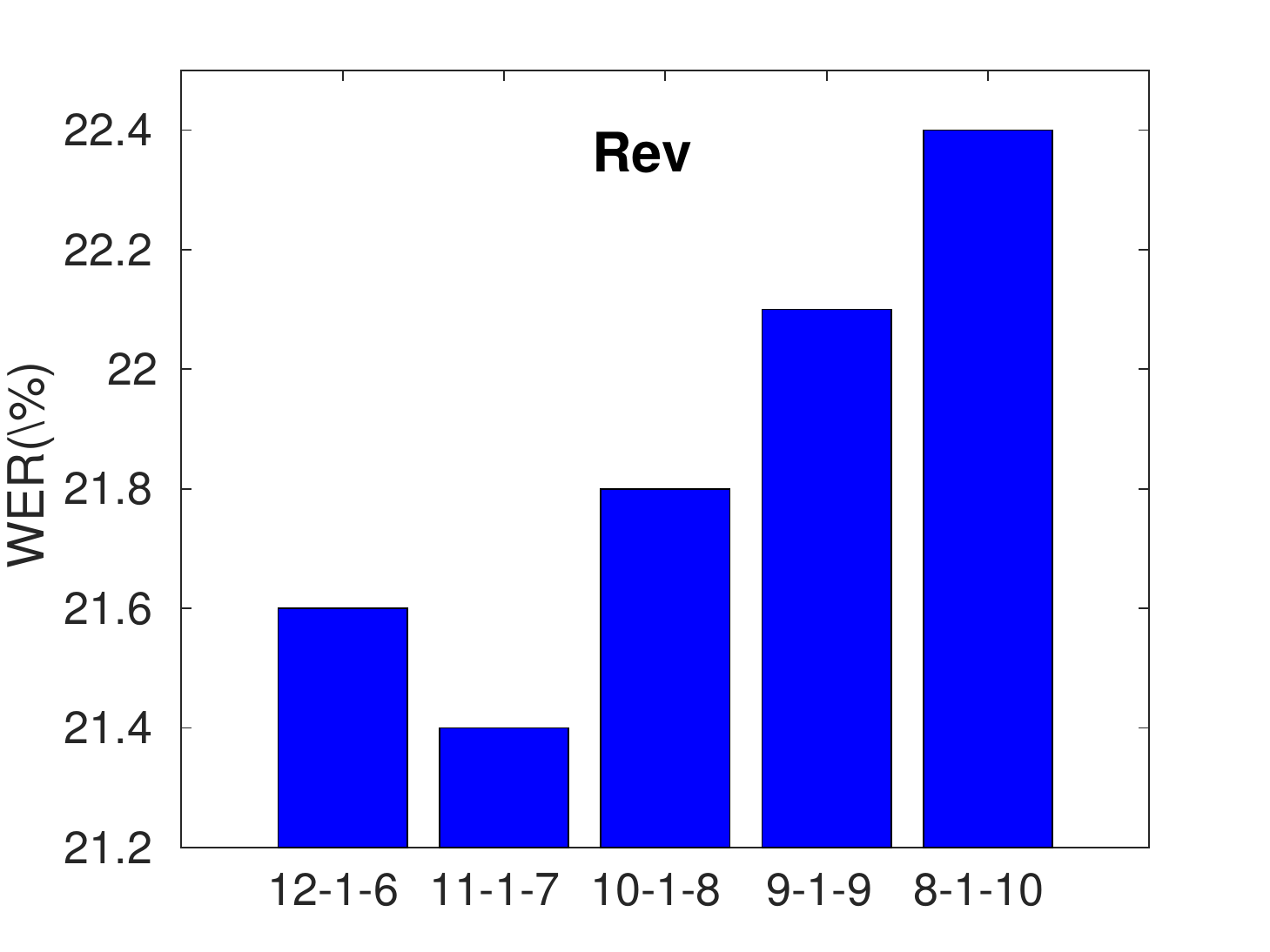}
\caption{Reverberated Scenario}
\label{fig:libri_rev}
\end{subfigure}
\caption{WER(\%) obtained with different context configurations for the close-talking and reverberated version of Librispeech (Test1, fmllr features).}
\label{fig:libri_res}
\end{figure}

In the previous experiment, we tested only fully asymmetric windows with $\rho_{cw}$=0\% (future frames) or $\rho_{cw}=100\%$ (past frames). However, it is worth addressing hybrid configurations, where both past and future frames are considered. With this purpose, Fig. \ref{fig:acw_scw} compares this kind of asymmetric window under both close-talking and distant-talking conditions, using contexts of different durations. 
For each CW length, the asymmetric CW curve represents the best ASR performance among all the configurations that derive from varying the balance factor $\rho_{cw}$.
Fig. \ref{fig:libri_res} shows the results obtained with the Librispeech task, by adopting the CW lengths that turned to be optimal in the case of DIRHA-WSJ task (i.e., 11 in the close-talking condition and 19 for the reverberated case).

%we can preliminarily suppose that an asymmetric context windows of about 19 frames composed of 11 past frames, the current frame, and 7 future frames (in the following denoted as 11-1-7) will be appropriate for the addressed distant-talking tasks. This kind of asymmetric window would have an asymmetricity index $\rho_{cw}$ of 61\%. %For close-talking speech recognition, standard symmetric context windows seems more adequate. In particular, one can note that in the later scenario a context window of 11 frames is optimal against a time context of 19 frames suitable for distant-talking cases. This means that a larger context helps the DNN in counteracting the uncertainty originated by challenging environmental conditions. 
%Based on this result, Fig. \ref{fig:acw_scw} compares asymmetric windows (based on $\rho_{cw}$ of 60-65\%) with symmetric windows for different context durations. %For ACW experiments, we employ a  $\rho_{cw}$  of about 60\%, which is chosen according to the results of Fig. \ref{fig:past_future_rev}.
 
%\begin{figure*}[t]
%\centering
%  \includegraphics[scale=0.65]{ACW_simrev.pdf}
%\caption{Comparison between symmetric and asymmetric windows of different durations in a reverberated distant-talking scenario (Rev) using fMLLR features computed on the DIRHA-WSJ-rev task.}\label{fig:acw_scw}
%\end{figure*}

From the close-talking experiments, it emerges that the standard SCW slightly outperforms the best asymmetric one, as clearly  highlighted by Fig. \ref{fig:acw_cl}. This trend is also confirmed in Fig. \ref{fig:libri_cl}, where different context windows of length 11 have been tested on the close-talking version of Librispeech. 
In both cases, the gap between symmetric and asymmetric contexts is not so large (on average less than 2\% relative decrease),  but it suggests to use an ACW in close-talking scenarios only when real-time/low-latency constraints arise. 

Differently, Fig. \ref{fig:acw_rev} shows that the asymmetric window consistently outperforms the standard symmetric one in the distant-talking case, for all the considered context durations. On average, about 5\% relative WER decrease is obtained with, essentially, no additional computational cost. This result is also confirmed by Fig. \ref{fig:libri_rev}, that reports the performance obtained on the reverberated Librispeech for various CW settings. This figure not only shows that an asymmetric window that embeds more past than future frames is a proper choice when reverberation arises, but it also highlights that the opposite setting (i.e., embedding more future than past frames) leads to a rather significant loss of performance.  

Previous experiments were based on fMLLR features. In Table \ref{tab:fea} we extend the experimental validation to other acoustic features, such as FBANK and MFCC coefficients. We also consider CNNs as an alternative to the fully-connected DNNs used so far. %For the following experiments, the best symmetric and asymmetric context windows obtained in the previous experiments are adopted.
\begin{table}[t!]
\centering
\tabcolsep=0.50cm
    \begin{tabular}{  | l | c | c | c | c | c |}
    \cline{1-4}
Architecture & Features & SCW (9-1-9) & ACW (11-1-7) \\ \hline
DNN & fMLLR & 15.2 & \textbf{14.8} \\ \hline
DNN & MFCC & 21.8 & \textbf{20.8} \\ \hline
DNN & FBANK & 20.7 & \textbf{20.2} \\ \hline
CNN & FBANK & 18.5 & \textbf{18.1 }\\ \hline

\end{tabular}
\caption{Comparison between the WERs(\%) achieved with SCWs and ACWs, when different features and DNN architectures are used.}
\label{tab:fea}
\end{table}
Results confirm that the ACW outperforms the symmetric one in all the considered settings. %Although the best performance is achieved with fMLLR features, the computation of such coefficients is not compliant with real-time constraints, since an additional decoding step performed with a HMM-GMM acoustic model is required. For real-time applications, standard MFCCs or FBANKs thus remain the most viable choice. 
The last row of Table \ref{tab:fea} also highlights an interesting performance improvement achieved with CNNs. CNNs
%that represent a valid alternative to fully-connected DNNs, 
are based on local connectivity, weight sharing, and pooling operations that allow them to exhibit some invariance to small feature shifts along the frequency axis, with well-known benefits against speaker and environment variations \cite{cnn1}. Hence, they represent a valid alternative to fully-connected DNNs, also jointly used with ACW under reverberant conditions. 

\begin{figure}[t!]
\begin{subfigure}{0.50\textwidth}
\includegraphics[scale=0.44]{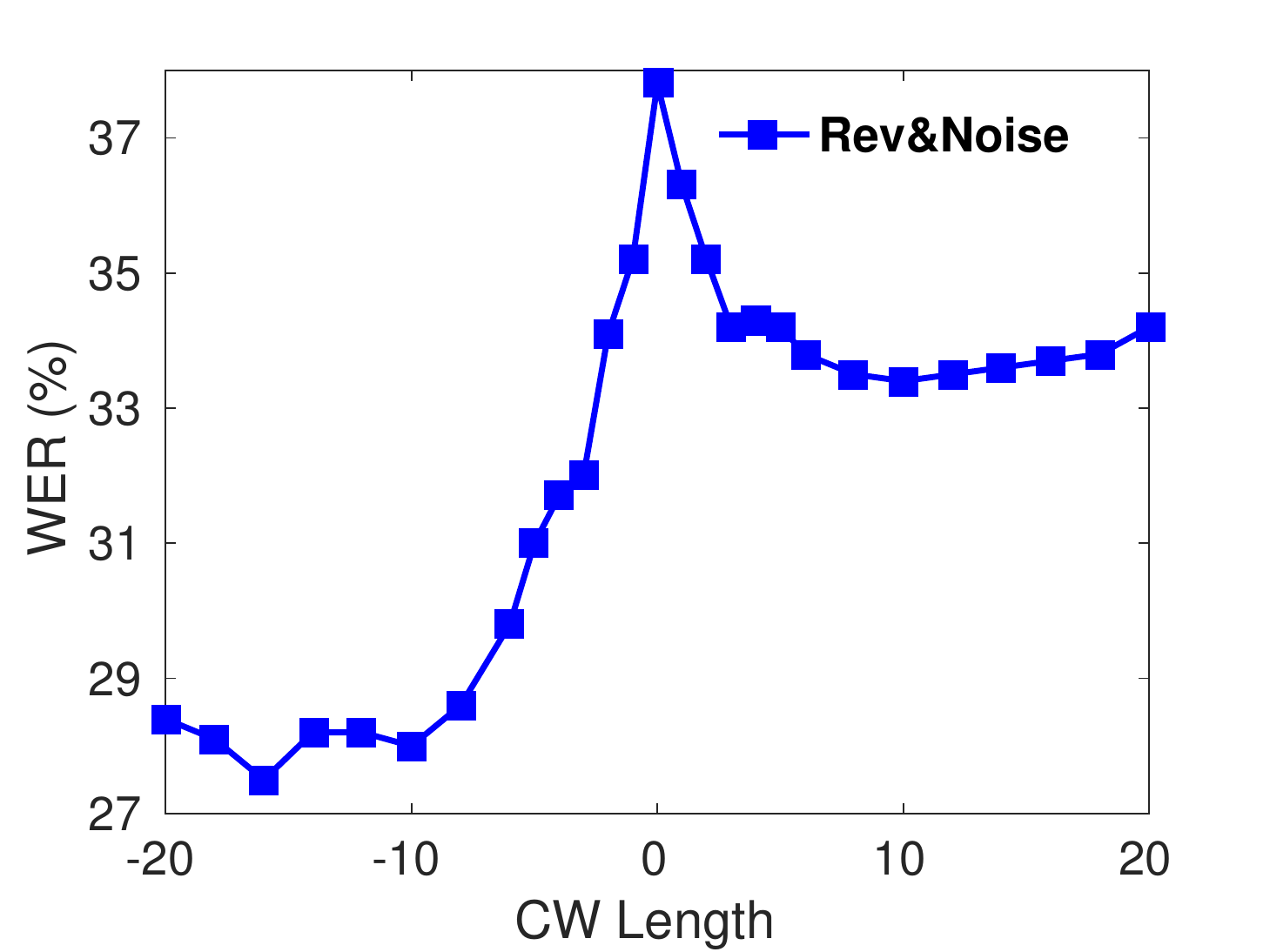}
\caption{Fully asymmetric context}
\label{fig:ab_revnoise}
\end{subfigure} \hspace{0.0\textwidth}
\begin{subfigure}{0.50\textwidth}
\includegraphics[scale=0.44]{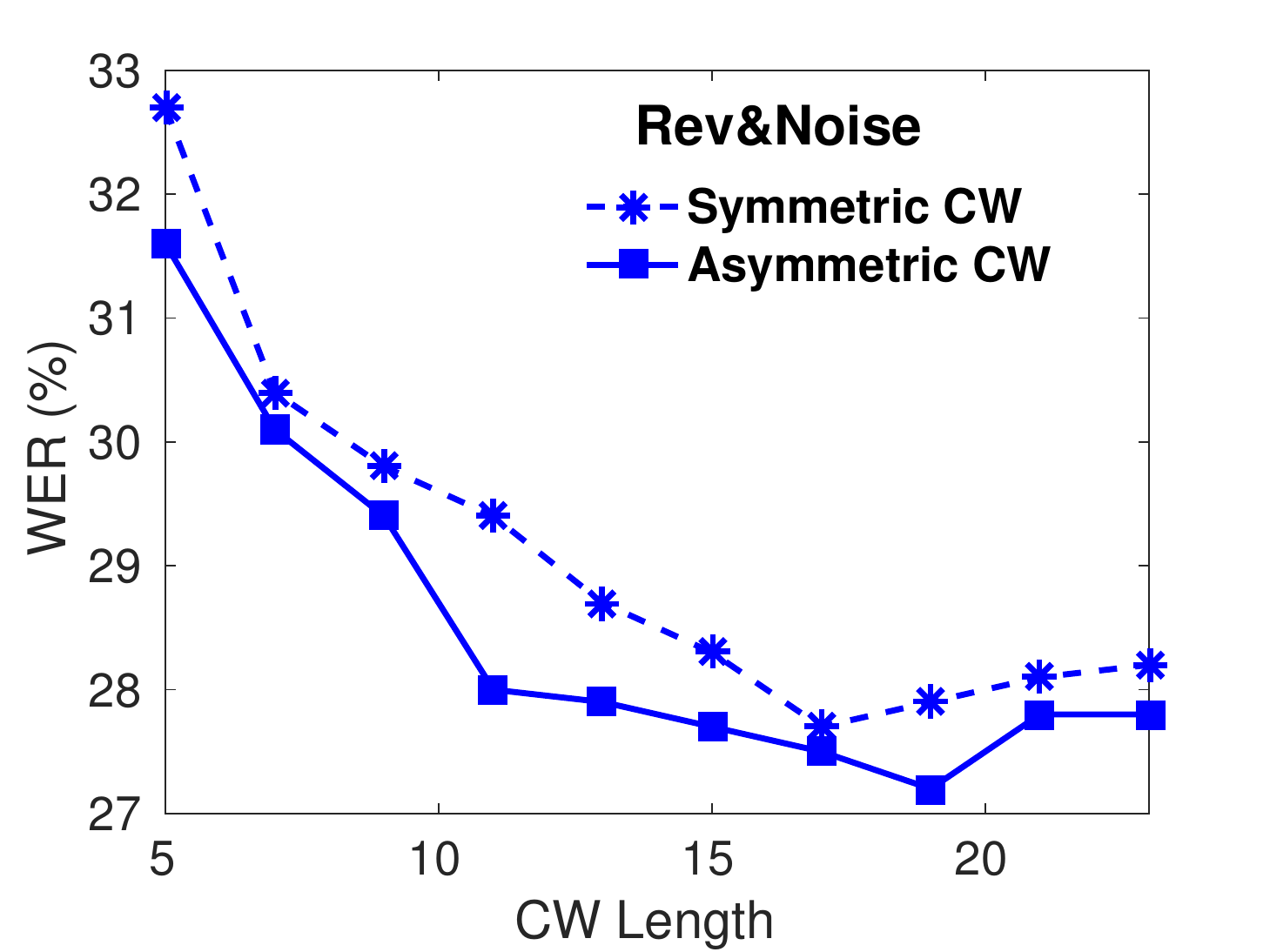}
\caption{Symmetric vs asymmetric CW.}
\label{fig:acw_revnoise}
\end{subfigure}
\caption{Comparison between SCW and ACW under mismatched conditions. Training is performed using reverberated data (using WSJ-rev), while test material is corrupted by both noise and reverberation (DIRHA-WSJ-rev\&noise). fMLLR features are used in this experiment.}
\label{fig:acw}
\end{figure}

To study the effectiveness of asymmetric contexts under mismatching conditions (that often arise in real applications), we now train the DSR system with reverberated data (Rev) and test it on real signals (DIRHA-WSJ-rev\&noise) affected  by both noise and reverberation.
Fig. \ref{fig:ab_revnoise} shows the results obtained when fully ACWs are adopted. Fig. \ref{fig:acw_revnoise}, instead, compares symmetric and optimal asymmetric windows, with different CW lengths.  
Due to the more challenging conditions characterizing this test, WER(\%) is significantly worse than that highlighted in Fig. \ref{fig:past_future_rev} and Fig. \ref{fig:acw_scw}. %, where matching conditions were considered. 
However, it is worth noting that the benefits deriving from the use of ACWs are maintained even under the addressed mismatching case.

%\subsection{ASR performance under mismatched conditions}
\subsection{ASR experiments with automatic context window composition}
\label{sec:mis}
As discussed in Sec. \ref{sec:algo}, the hyperparameters $N_{p}$ and $N_{f}$ of the ACW can be derived by applying $AutoCW$ (Alg. 1).
In this section, we conduct a set of experiments to evaluate the loss of performance introduced by it, when compared to the ideal (and computational expensive) conditions under which previous experiments (in Fig. \ref{fig:acw_revnoise}) were performed. Let us recall that, in the latter cases, a grid optimization was done over all the possible CW combinations.

%and that the best performance, 27.2 \% WER, was obtained using an optimal CW 11-1-7, that is an overall length of 19 frames. 

The first row of Table \ref{tab:last_res} shows the results obtained with the aforementioned mismatching condition. The best performance, 27.2 \% WER, is obtained using an optimal CW 11-1-7, that is an overall length of 19 frames.
It is however worth noting that applying $AutoCW$ leads to a very similar combination, i.e., 12-1-6, is obtained, which corresponds to a comparable recognition performance, i.e., 27.3\% WER.
The last two rows, instead,  report the results achieved with the reverberated version of Librispeech.
In this case, the proposed $AutoCW$ algorithm provides a CW setting that corresponds to the optimal choice. 
For both recognition tasks, we can also observe that applying $AutoCW$ leads to 2-3\% relative reduction of WER provided by the SCW.

%...The last column of Tab. \ref{tab:time} reports the results obtained with the proposed automatic method, where instead of performing a grid optimization over all the various context window combinations, we exploit the gradient norm analysis.
%Interestingly, the algorithm is able to select a nearly optimal $\rho_{cw}$ for all the different reverberation times. For the close-talking scenario ($T_{60}$=0 ms),  a SCW is selected, while for distant-talking experiments ACWs that embed more past that future frames are composed. %Similarly to the optimal case, the asymmetricity factor $\rho_{cw}$ retrieved by the automatic procedure tends to increase as the reverberation time increases. 
%To extend the latter results to real recordings, the first row of Tab. \ref{tab:last_res} shows the performance obtained with the DIRHA-WSJ rev\&noise recordings. %The last two rows, instead,  report the results achieved with the reverberated version of Librispeech.
%In all these cases, the proposed algorithm still provides a context window setting that is nearly optimal. ...

\begin{table}[t!]
\centering
\tabcolsep=0.30cm
    \begin{tabular}{  | l | c | c | c | c | c |}
    \cline{1-5}
\multicolumn{2}{ | c | }{ $T_{60}$(ms)} & SCW & ACW (opt) & AutoCW\\ \hline
\multirow{2}{*}{DIRHA-WSJ-rev\&noise} & CW & \textit{ 8-1-8} & \textit{11-1-7} & \textit{12-1-6} \\  \cline{2-5}
                  & WER    & 27.9 & \textbf{27.2} & 27.3 \\ \hline
                   
\multirow{2}{*}{Rev-LibriSpeech (Test1)} & CW & \textit{ 9-1-9} & \textit{11-1-7} & \textit{11-1-7} \\  \cline{2-5}
                   & WER   & 22.1 & \textbf{21.4} & 21.4 \\ \hline
           
 \multirow{2}{*}{Rev-LibriSpeech (Test2)} & CW & \textit{9-1-9} & \textit{11-1-7} & \textit{11-1-7} \\  \cline{2-5}
                    & WER  & 51.3 & \textbf{50.1} & 50.1 \\ \hline
                      
    \end{tabular}
\caption{Comparison between WER(\%) obtained with SCW, the optimal asymmetric one (ACW opt), and with the context configuration derived by our algorithm ($AutoCW$). The experiments are performed with fMLLR features.} 
\label{tab:last_res}
\end{table}

%In real applications, DSR systems often operate under a mismatched condition.
%In practice, despite best efforts to avoid this issue, test is normally characterized by noise and reverberation different from those actually faced during training. 
%It is thus of interest to study the effectiveness of the proposed method even a certain degree of mismatch occurs.

%Similarly to Fig. \ref{fig:past_future_rev}, negative x-axis represents past frames, while positive x-axis refers to future frames contexts.  
% \begin{figure}
% \begin{minipage}{0.55\linewidth}
%   \centering
%    \includegraphics[width=\linewidth]{after_before_revnoise.pdf}
%   \label{fig:blah1}
% \end{minipage}
% \begin{minipage}{0.55\linewidth}
%   \centering
%    \includegraphics[width=\linewidth]{ACW_realrevnoise.pdf}
%   \label{fig:blah2}
% \end{minipage}
% \caption{Comparison between the WER(\%) performance obtain with the symmetric and the proposed asymmetric context window in mismatching conditions.}\label{fig:revnoise}
% \end{figure}

%Interestingly enough,  the asymmetric trend previously observed in Sec. \ref{sec:acw_exp}, is maintained even under mismatching conditions. %Similarly to Sec. \ref{sec:acw_exp}, this result suggest that an asymmetric context gathering 11 past and 7 future frames ($\rho_{cw}=61\%$) would be adequate for also the addressed conditions

Another set of experiments concerns a different kind of mismatch that occurs when training and test are performed in different acoustic environments.
As reported in Table \ref{tab:mism}, training was carried out in the DIRHA living-room, using the WSJ-rev corpus, while test is performed in three different contexts, i.e., an office, a surgery room, as well as a room of another apartment. The test data were generated following the approach described in Sec. \ref{sec:dt_data} for DIRHA-WSJ-rev, but using real IRs that were measured in the aforementioned environments. %For these experiments, we inherit a context length of 19 frames, that was optimal in the previous study. %In particular, the adopted context windows embeds 11 past and 7 future frames, while the symmetric one is based on 9 past and 9 future frames.
\begin{table}[t!]
\centering
\tabcolsep=0.25cm
    \begin{tabular}{  | l | c | c | c | c | c |}
    \cline{1-4}
\multirow{2}{*}{{\backslashbox{\em{Env.}}{\em{Context Wind.}}}}  & SCW (opt) & ACW (opt) & AutoCW\\     \cline{2-4}
 & 9-1-9 & 11-1-7 & 12-1-6\\ \hline
Office ($T_{60}=650$ ms) & 16.6 & \textbf{16.2} & 16.2 \\ \hline
Home ($T_{60}=700$ ms)  & 19.5 & \textbf{19.1} & 19.3 \\ \hline
Surgery Room ($T_{60}=850$ ms) & 21.4 & \textbf{20.3} & 20.5 \\ \hline
\end{tabular}
\caption{WER(\%) obtained with SCWs and ACWs, in different acoustic environments and under mismatched conditions. Training is performed in the DIRHA livingroom ($T_{60}$=750ms) using WSJ-rev, while test is performed in different acoustic environments with different reverberation times.}
\label{tab:mism}
\end{table}
Results show that the use of ACW introduces advantages in terms of ASR performance under all the tested conditions, even when training and test are performed in different acoustic environments. Moreover, the application of $AutoCW$ provides a CW composition 12-1-6, very similar to the optimal one, which corresponds to a 2-3\% relative WER reduction, if compared to the performance obtained using SCW.

\subsection{Performance analysis with different reverberation times}
\label{sec:t60}
%In the previous experiments, training was performed in the targeted domestic environment. 
As pointed out above, the application of $AutoCW$ can have a different impact according to the reverberant conditions under which training and test are performed.
Concerning this, we further extended our validation by simulating acoustic environments with increasing reverberation times $T_{60}$. For this study, a set of IRs simulated with the image method \cite{image} were used to contaminate both training (WSJ-clean) and testing corpora (DIRHA-WSJ-clean). Table \ref{tab:time} summarizes the results obtained with $T_{60}$ ranging from 250 ms to 1000 ms. 

\begin{table}[t!]
\centering
\tabcolsep=0.40cm
    \begin{tabular}{  | l | c | c | c | c | c |}
    \cline{1-5}
\multicolumn{2}{ | c | }{ $T_{60}$(ms)} & SCW & ACW (opt)  & AutoCW\\ \hline
\multirow{2}{*}{0 ms} & CW & \textit{ 5-1-5} & \textit{6-1-4} &  \textit{5-1-5}\\  \cline{2-5}
                  & WER    & \textbf{3.6} & 3.7  & 3.6 \\ \hline
                   
\multirow{2}{*}{250 ms} & CW & \textit{ 5-1-5} & \textit{6-1-4} & \textit{6-1-4} \\  \cline{2-5}
                   & WER   & 5.5 & \textbf{5.1} & 5.1 \\ \hline
           
 \multirow{2}{*}{500 ms} & CW & \textit{6-1-6} & \textit{7-1-5}  & \textit{8-1-4} \\  \cline{2-5}
                    & WER  & 9.1 & \textbf{8.5} & 8.7 \\ \hline

 \multirow{2}{*}{750 ms} & CW  & \textit{9-1-9} & \textit{12-1-6}  & \textit{11-1-7}\\  \cline{2-5}
                    & WER  & 15.2 & \textbf{14.8}  & 14.9 \\ \hline
                      
 \multirow{2}{*}{1000 ms} & CW  & \textit{12-1-12} & \textit{18-1-6}  & \textit{19-1-5}\\  \cline{2-5}
                    & WER  & 20.5 & \textbf{20.1} & 20.1 \\ \hline
                      
    \end{tabular}
\caption{Comparison between WER(\%) obtained with SCW and ACW under different reverberation conditions. The last column reports the results obtained with the proposed $AutoCW$ algorithm.}
\label{tab:time}
\end{table}

\begin{figure}[t!]
\begin{subfigure}{0.50\textwidth}
\includegraphics[scale=0.44]{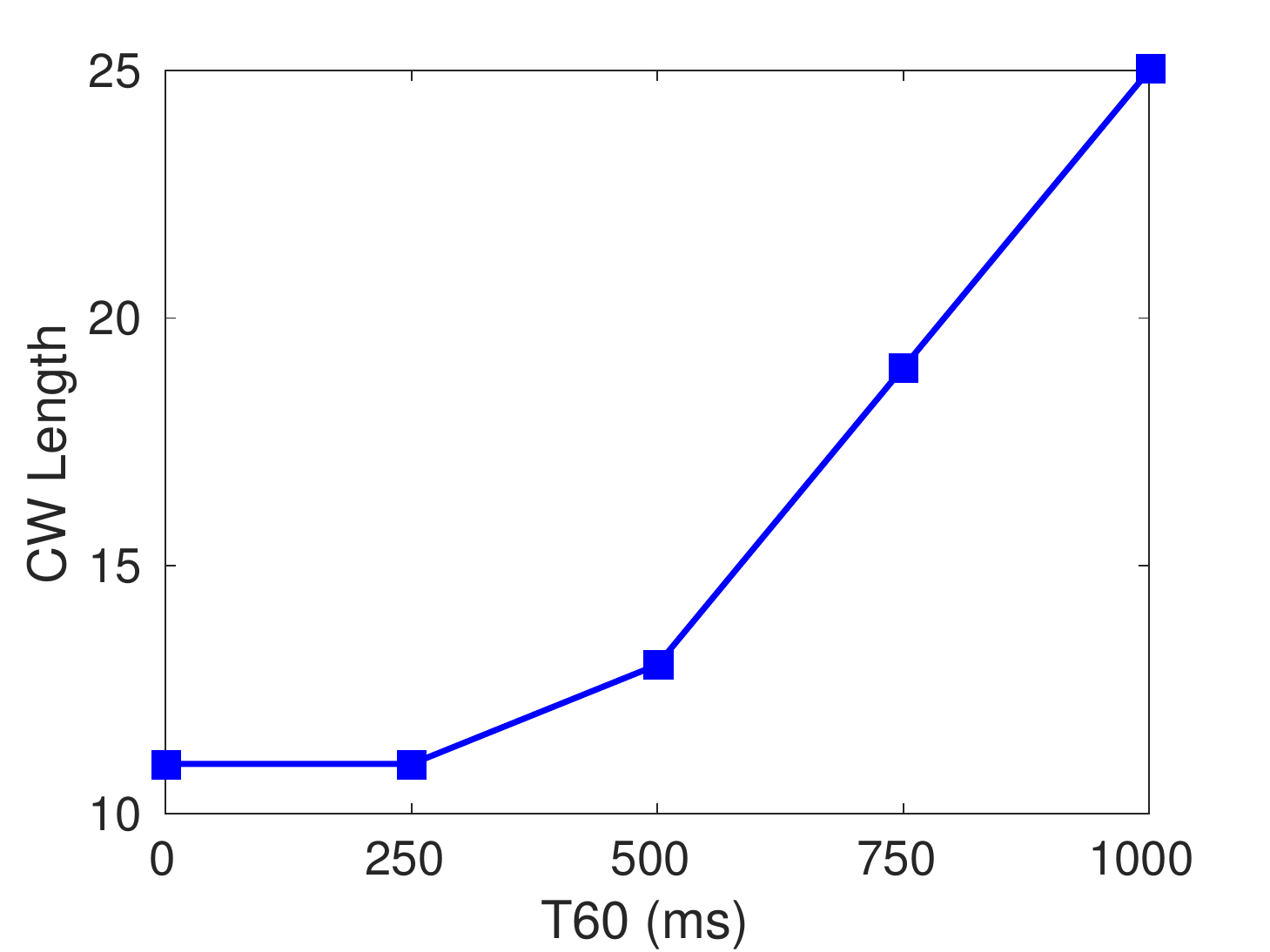}
\caption{Length of the context window vs $T_{60}$.}
\label{fig:cwlent60}
\end{subfigure} \hspace{0.0\textwidth}
\begin{subfigure}{0.50\textwidth}
\includegraphics[scale=0.44]{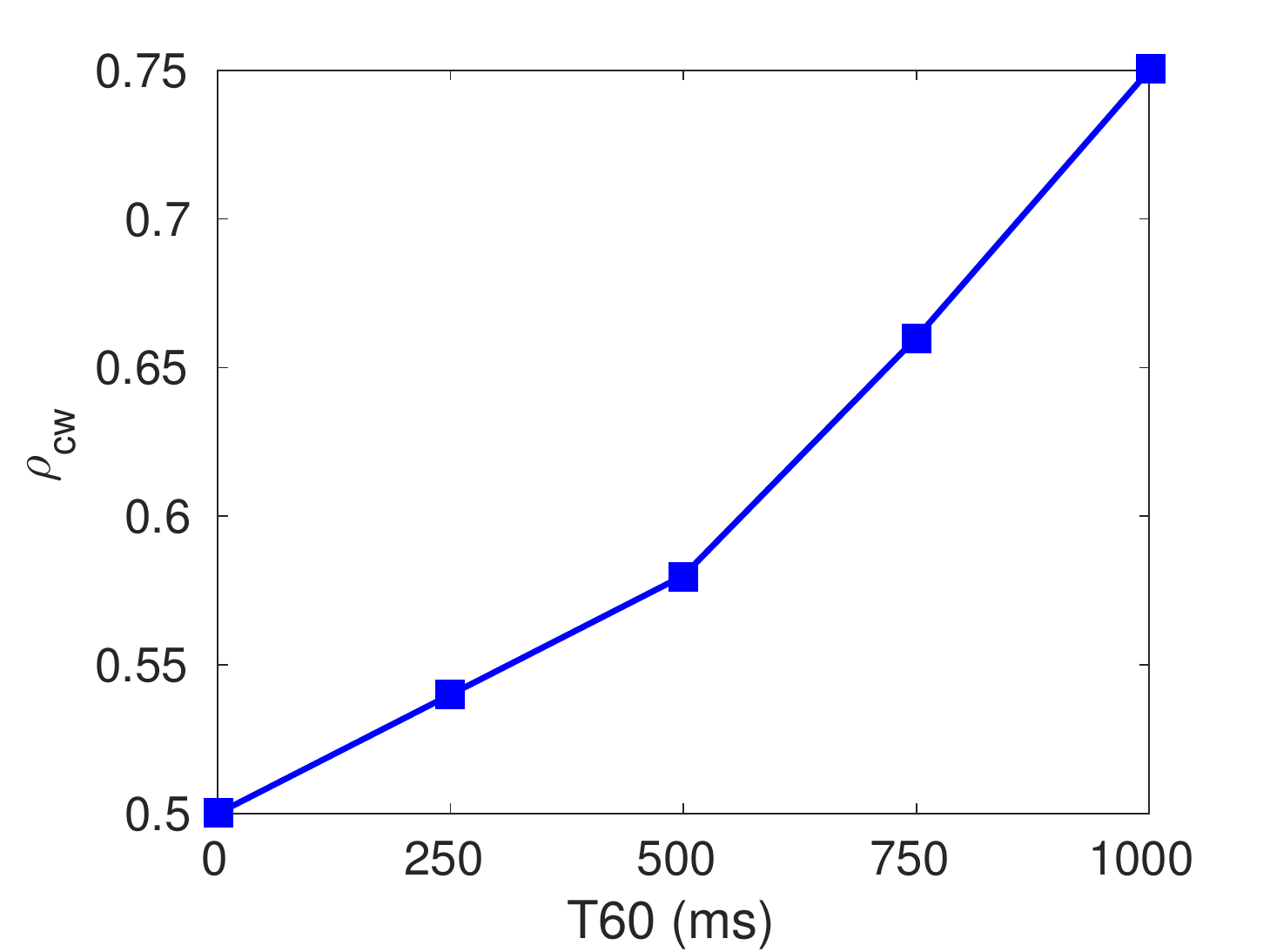}
\caption{$\rho_{cw}$ vs $T_{60}$.}
\label{fig:rhot60}
\end{subfigure}
\caption{Main features of the optimal context window for different reverberation times.}
\label{fig:t60_pict}
\end{figure}

As expected, results show that the performance progressively degrades as $T_{60}$ increases. More interestingly, the asymmetric window is able to overtake standard symmetric ones in all the explored reverberant conditions. It is also worth noting that
%as highlighted in Fig. \ref{fig:cwlent60}, 
larger contexts are needed when increasing the reverberation time, as highlighted in Fig. \ref{fig:cwlent60}.  
For instance, when $T_{60}$=250 ms the optimal window integrates only 11 frames, while 25 frames are necessary when $T_{60}$=1000 ms. Interestingly enough, the coefficient $\rho_{cw}$, that measures the amount of asymmetricity in the CW, increases as the reverberation time increases (see Fig. \ref{fig:rhot60}). This means that 
%for large $T_{60}$s, 
the reverberation effects significantly 
%corrupt
reduce the usefulness of
future frames in the case of large $T_{60}$s, which makes convenient the use of more asymmetric context windows.
It is worth noting that the proposed $AutoCW$ algorithm provides nearly optimal contexts, that lead to a negligible performance reduction over the best CW for all the considered reverberation times.  
Under close-talking conditions ($T_{60}$=0 ms), $AutoCW$ correctly derives a symmetric context window of 11 frames. Similarly to the optimal case, the proposed method correctly provides longer and more asymmetric context windows when reverberation increases.

%As observed in the previous section, the $\rho_{cw}$ factor is a function of the reverberation time. For the DIRHA-English experiments, the best context window guessed by the proposed algorithm is $12-1-6$ ($\rho_{cw}$=0.66). From Fig. \ref{fig:rhot60}  one can observe that this value of $\rho_{cw}$ corresponds to a $T_{60}$ of about $750$ ms, which is approximately the average reverberation time measured during the real recordings of the DIRHA-English dataset. The proposed algorithm can thus also be used to provide a rough estimate of the $T_{60}$.

\section{Conclusions}
\label{sec:conc}
In this paper, we extensively studied the role played by ACWs to counteract the adverse effects of reverberation in a distant speech recognizer. Under these environmental conditions, this windowing mechanism has proven to be a viable alternative to a more standard symmetric context. The asymmetric window, in fact, feeds the DNN with a more convenient frame configuration which carries, on average,  information that is less redundant and less affected by the correlation effects introduced by reverberation.

To optimize the characteristics of the asymmetric context window, this work proposed a novel algorithm that analyzes the norm of the DNN gradients over the various input frames.  The $AutoCW$ algorithm, tested on different tasks, datasets, and environments turned out to derive nearly optimal windows under different acoustic conditions. Our method, that is characterized by a linear computational complexity, is significantly more efficient  than  the traditional grid search optimization over all the possible frame configurations, which has a quadratic complexity.

As previously mentioned, an open issue is represented by the flexibility of the proposed approach to tackle possible changes of the reverberant conditions. This issue can be addressed in several possible ways, for instance by combining the current solution with a pre-processing step that realizes a preliminary environmental classification, which aims at selecting in real-time the most suitable asymmetric context as well as the related neural network.

Overall, the use of ACW and of $AutoCW$ turns out to represent a simple and effective approach to improve DSR performance under noisy and reverberant conditions, in particular with medium-high reverberation times, and for the development of real-time low-complexity applications.

%This work provided experimental evidence at different stages, reporting results at signal, feature, gradient as well as ASR performance levels. The ASR performance has been evaluated on different window configurations, DNN architectures, and acoustic conditions. 
%With the purpose of optimizing the two hyperparameters of the ACW with a linear computational complexity rather than with a quadratic grid search, we also proposed a simple algorithm based on a gradient analysis of the input frames. 

%The proposed algorithm turned out to be effective for automatically composing the various frames of the context window, leading to performance improvement over a standard symmetric frame configuration. This improvement is obtained without introducing additional computational costs, making ACW particularly suitable for small-footprint and on-line DSR systems.

\section*{References}

\bibliography{mybibfile}

\end{document}